\documentclass[10pt,journal,compsoc]{IEEEtran}
\usepackage{xcolor}

\usepackage{amsfonts, bm}
\newcommand\eat[1]{}
\usepackage[colorlinks,linkcolor={blue!50!black},citecolor={blue!50!black},urlcolor={blue!50!black}]{hyperref}
\newcommand{\C}[1]{#1}  %

\def\br{{\mathbf r}}

\def\bv{{\mathbf v}}
\def\w{{\mathbf w}}
\def\x{{\mathbf x}}
\def\y{{\mathbf y}}
\def\z{{\mathbf z}}

\def\hatr{{\mathbf{\hat r}}}

\def\hatx{{\mathbf{\hat w}}}
\def\hatx{{\mathbf{\hat x}}}
\def\haty{{\mathbf{\hat y}}}

\def\hatm{{\mathbf{\hat m}}}

\def\barv{{\mathbf{\bar v}}}
\def\barw{{\mathbf{\bar w}}}
\def\barx{{\mathbf{\bar x}}}
\def\bary{{\mathbf{\bar y}}}
\def\barz{{\mathbf{\bar z}}}

\usepackage{mathtools}
\usepackage{amsmath}
\usepackage{subcaption}
\usepackage{makecell}
\usepackage{tabularx}
\usepackage[ruled,vlined]{algorithm2e}
\usepackage{pifont}
\usepackage{multirow}
\usepackage{makecell}
\newcommand{\cmark}{\ding{51}}
\newcommand{\xmark}{\ding{55}}

\DeclarePairedDelimiter\round{\lfloor}{\rceil}

\usepackage{amsmath,amsfonts,bm}

\def\eqref#1{equation~\ref{#1}}

\def\1{\bm{1}}

\DeclareMathAlphabet{\mathsfit}{\encodingdefault}{\sfdefault}{m}{sl}
\SetMathAlphabet{\mathsfit}{bold}{\encodingdefault}{\sfdefault}{bx}{n}

\ifCLASSOPTIONcompsoc
  \usepackage[nocompress]{cite}
\else
  \usepackage{cite}
\fi

\ifCLASSINFOpdf
\else
\fi

\begin{document}
\title{Insights from Generative Modeling \\ for Neural Video Compression}
\author{Ruihan~Yang,
        Yibo~Yang,
        Joseph~Marino,
        and~Stephan~Mandt,~\IEEEmembership{Member,~IEEE}
\IEEEcompsocitemizethanks{\IEEEcompsocthanksitem Ruihan~Yang,~Yibo~Yang~and~Stephan~Mandt are with the Department
of Computer Science, University of California Irvine.\protect
\IEEEcompsocthanksitem Joseph Marino is with DeepMind, formerly with Caltech.
}%
}

\IEEEtitleabstractindextext{%
\begin{abstract}
While recent machine learning research has revealed connections between deep generative models such as VAEs and rate-distortion losses used in learned compression, most of this work has focused on images. In a similar spirit, we view recently proposed neural video coding algorithms through the lens of deep autoregressive and latent variable modeling. We present these codecs as instances of a generalized stochastic temporal autoregressive transform, and propose new avenues for further improvements inspired by normalizing flows and structured priors. We propose several architectures that yield state-of-the-art video compression performance on high-resolution video and discuss their tradeoffs and ablations. In particular, we propose (i) improved temporal autoregressive transforms, (ii) improved entropy models with structured and temporal dependencies, and (iii) variable bitrate versions of our algorithms. Since our improvements are compatible with a large class of existing models, we provide further evidence that the generative modeling viewpoint can advance the neural video coding field.

\end{abstract}

}

\maketitle

\IEEEdisplaynontitleabstractindextext

\IEEEpeerreviewmaketitle

\IEEEraisesectionheading{\section{Introduction}\label{sec:introduction}}

\IEEEPARstart{N}{eural} data compression \cite{yang2022introduction} has evolved to become a promising new application and testing domain for generative modeling\footnote{Proceedings of the First ICLR Workshop on Neural Compression: From Information Theory to Applications, \url{neuralcompression.github.io}}. Generative models such as hierarchical variational autoencoders have already demonstrated empirical improvements in image compression, outperforming classical codecs \cite{minnen2018joint, yang2020improving} such as BPG \cite{bellard2014}. For neural video compression, progress has proved harder due to complex temporal dependencies operating at multiple scales. Nevertheless, recent neural video codecs have shown promising performance gains \cite{agustsson2020scale}, in some cases on par with current hand-designed, classical codecs, e.g., HEVC\cite{wiegand2003overview, sullivan2012overview}. Compared to hand-designed codecs, learnable codecs are not limited to a specific data modality and offer a promising approach for streaming specialized content, such as sports or video conferencing. Therefore, improving neural video compression is vital for dealing with the ever-growing amount of video content being created.
    
The common approach to source compression transforms the data into a white noise distribution that can be more easily modeled and compressed with an entropy model. This way, data compression fundamentally involves decorrelation. Improving a model's capability to decorrelate data helps improves its compression performance. On the other hand, we can improve the entropy model (i.e., the model's prior) to capture any remaining dependencies. Just as compression techniques attempt to \textit{remove} structure, generative models attempt to \textit{generate} structure. For example, autoregressive flows map between less structured distributions, e.g., uncorrelated noise, and more structured distributions, e.g., images or video \cite{dinh2014nice,dinh2016density}. The inverse flow can remove dependencies in the data, making it more amenable to compression. Thus, a natural question to ask is how autoregressive flows can best be utilized in compression and if mechanisms in existing compression schemes can be interpreted as normalizing flows. 
    
This paper draws on recent insights in hierarchical sequential latent variable models with autoregressive flows \cite{marino2020improving}. In particular, we identify connections between this family of models and recent neural video codecs based on motion estimation \cite{lu2019dvc, liu2019learned, lin2020m, lu2020content, agustsson2020scale,yang2020autoregressive}.
By interpreting this technique as an instantiation of a more general autoregressive flow transform, we propose various alternatives and improvements based on insights from generative modeling. Specifically, our contributions are as follows:

First, we interpret existing video compression methods through the framework of generative modeling and variational inference, allowing us to readily investigate extensions and ablations. 
\C{In particular, we discuss the relationship between sequence modeling and sequence compression. We identify autoregressive transform as a key component in both cases and suggest incorporating it in a sequential latent variable as the basis of our approach.}

Second, we improve a popular model for neural video compression, Scale-Space Flow (SSF)~\cite{agustsson2020scale}. This model uses motion estimation to predict the frame being compressed and further compresses the residual obtained by subtraction. Our proposed model extends the SSF model with a more flexible decoder and prior, and 
obtains improved rate-distortion performance.
Specifically, we incorporate a learnable scaling transform to allow for more expressive and accurate reconstruction. Augmenting a shift transform by scale-then-shift is inspired by the extension of NICE \cite{dinh2014nice} to RealNVP \cite{dinh2016density}. 
We also compare motion-estimation-based vs.~purely CNN-based approaches to predictive coding.

Third, our probabilistic perspective allows us to explore improved entropy models for SSF and its relatives. In particular, we explore structured priors that jointly encode motion and residual information. As the two tend to be spatially correlated, encoding residual information conditioned on motion information results in a better prior. Since the residual dominates the bitrate, our improved entropy model reduces the overall bitrate significantly.
We also investigate different versions of temporal priors, where we either condition on latent variables or on frame reconstructions and discuss their tradeoffs in terms of bit savings and optimization challenges. 

\C{Finally}, also from the perspective of generative modeling, we present variable bitrate versions of our models, i.e., training a single encoder and decoder that works at different points on the rate-distortion curves. This step is considered important for making neural coding schemes practical. Our experiments show that the difference in rate-distortion performance of variable bitrate models and models tuned to individual bit rates depends on the model complexity.

Our paper is structured as follows. We establish our viewpoint and model improvements in Section~\ref{sec:method-journal}, discuss related work in Section~\ref{sec:related-work}, and present experiments in Section~\ref{sec:experiments}. Conclusions are provided in Section~\ref{sec: discussion}.

\section{Video Compression through Deep Autoregressive Modeling}
\label{sec:method-journal}
We first review low-latency video compression, including the classical predictive coding technique.
We then draw connections between data compression and data modeling with a Masked Autoregressive Flow (MAF) \cite{papamakarios2017masked}, a generative model based on a temporal autoregressive transform that resembles predictive coding. Finally, inspired by hierarchical autoregressive flow models \cite{marino2020improving}, we combine the strength of autoregressive modeling with the end-to-end optimizable transform coding approach of VAEs \cite{balle2020nonlinear}, resulting in a sequential VAE \cite{chung2015recurrent} with an autoregressive encoding/decoding transform. The resulting model captures many existing neural video compression methods \cite{lu2019dvc, liu2019learned, lin2020m, lu2020content, agustsson2020scale,yang2020autoregressive}, and serves as the basis of our proposed improvements to the decoding transform as well as the prior model.

\textbf{Notation.} We use bold letters (e.g., $\x, \z$) to denote random variables and variables with superscripts to indicate deterministically computed quantities (e.g., $\hatx, \barz$). We use $p(\x)$ to denote the probability distribution or density induced by random variable $\x$ and write $P(\x)$ to emphasize when it is a probability mass function.

\subsection{Background}\label{sec:background}
As follows, we review lossy video compression, focusing on the low-latency online compression setting. We then review the classical technique of predictive coding, which provides the high-level algorithmic framework of many video compression methods, including ours. As our final building block, we review normalizing flow models, in particular the Masked Autoregressive Flow for sequence modeling.

\textbf{Video Compression.} 
Given a typical sequence of video frames $\x_{1:T}$, 
lossy video compression ultimately aims to find a short bitstring description of $\x_{1:T}$, from which a faithful (but not necessarily perfect) reconstruction $\hatx_{1:T}$ can be recovered. Denoting the description length (``rate'') by $\mathcal{R}$ and the reconstruction error (``distortion'', often the Mean-Squared Error) by $\mathcal{D}$, lossy video compression aims to minimize the rate-distortion (R-D) objective function,
\begin{align}
    \mathcal{L} = \mathbb{E}_{\x_{1:T} \sim p_\text{data}}[ \mathcal{D}(\x_{1:T}, \hatx_{1:T}) + \beta \mathcal{R}(\hatx_{1:T}) ],
\end{align}
where $\beta > 0$ is a hyperparameter weighing the two costs, and the expectation is with respect to the source data distribution $p_{data}$, which is approximated by averaging over a training set of videos in practice. 
One simple approach is to compress each frame $\x_t$ separately using an image compression algorithm, which can exploit the spatial redundancy between the pixels within each frame. However, a key feature distinguishing video from image compression is the significant amount of temporal redundancy \emph{between} frames that can be exploited to improve the compression rate.

\textbf{Online Video Compression by Predictive Coding.}  In theory, the optimal rate-distortion performance is achieved by compressing exceedingly long blocks of frames together \cite{cover1999elements}. However, such an approach is rarely implemented in practice because of its prohibitive computation expense and the latency caused by buffering the frames into long blocks.
We consider video compression in the sequential/online setting, widely used in both conventional and recent neural codecs \cite{Rippel2019LearnedVC, agustsson2020scale} and are suitable for real-time applications such as video conferencing and live streaming.
In this setting, each video frame $\x_t$ is compressed in temporal order so that at any time $t < T$, the source frames up to time $t$, i.e., $\x_{1:t}$, are encoded and transmitted, and similarly the reconstructed frames up to time $t$, $\hatx_{1:t}$, are available to the receiver.
Since past frames are often highly indicative of future frames,
the basic idea for exploiting this temporal redundancy is to use knowledge of the previous frame reconstructions, $\hatx_{<t}$, to aid the compression of the current frame $\x_t$. Indeed, the earliest video codecs are based on transmitting frame differences $\x_t  - \hatx_{t-1}$ \cite{sullivan1998rate}, analogous to the classic modeling technique in dynamical systems whereby the state-space becomes first-order Markov when redefined in terms of temporal changes \cite{kalman1960new, marino2020improving}. 
This technique is further refined by \emph{predictive coding} \cite{cutler1952differential}, where instead of simply using the previous reconstructed frame $\hatx_{t-1}$, a \emph{motion-compensated prediction} of the current frame, $\barx_t$, is computed, and the residual $\x_t - \barx_t$ is transmitted instead.

In more detail, the idea of predictive coding is typically implemented in traditional video codecs as follows: 
\textbf{1. motion estimation}: The encoder estimates the motion vector $\mathbf{m}_t$ between the current frame $\x_t$ and the previous reconstruction $\hatx_{t-1}$; 
\textbf{2. motion compensation}: The encoder simulates the coding of $\mathbf{m}_t$, and uses the reconstruction $\hatm_t$ (as would be received by the decoder in step 3)
to transform the previous frame reconstruction $\hatx_{t-1}$ into a prediction of the current frame $\barx_t$. Conceptually, this is done by shifting the pixels of $\hatx_{t-1}$ according to the motion vector $\hatm_t$. The compression of $\mathbf{m}_t$ could either be lossy ($\hatm_t \approx \mathbf{m}_t$) or lossless ($\hatm_t = \mathbf{m}_t$).  
\textbf{3. residual compression:} The residual is computed as $\br_t = \x_t - \barx_t$, and is encoded. 
The decoder, upon receiving the bit-streams for $(\mathbf{m}_t, \br_t)$ and reconstructing them as $(\hatm_t, \hatr_t)$, computes the prediction $\barx_t$ as in step 2, %
and finally the reconstruction of the current frame, $\hatx_t = \barx_t + \hatr_t $.
The current reconstruction $\hatx_t$ then serves as the reference frame in the predictive coding of the next frame $\x_{t+1}$.

\textbf{Masked Autoregressive Flow (MAF).}
MAF \footnote{As a clarification, even though the original MAF was implemented with the masking approach of MADEs \cite{papamakarios2017masked} (hence ``masked'' in the name), we use the term ``MAF'' to refer more broadly to the normalizing flow defined by the temporal autoregressive transform in Eq.~\ref{eq:MAF}. }  \cite{papamakarios2017masked} is a type of normalizing flow \cite{papamakarios2021normalizing} that models the distribution of a random sequence $p(\x_{1:T})$ in terms of a simpler distribution $p(\y_{1:T})$ of its underlying noise variables $\y_{1:T}$. 
The two variables are related by the following invertible autoregressive transform,
\begin{align}
    \y_t &= \textstyle{\frac{\x_t - h_\mu(\x_{<t})} {h_\sigma(\x_{<t})}} \; \Leftrightarrow \; \x_t = h_\mu(\x_{<t}) + h_\sigma(\x_{<t}) \odot \y_t
    , \label{eq:MAF}
\end{align}
for $t= 1, 2, ..., T$. Here, $\odot$ denotes element-wise multiplication, $\x_{<t}$ denotes all frames up to time $t$, and $h_\mu$ and $h_\sigma$ are two deterministic neural network mappings \footnote{In general, $(h_\mu, h_\sigma)$ can be a different pair of neural networks for each time step $t$, although for practical sequence modeling, they are often shared across time and only receive a fixed length-$k$ context $x_{(t-k+1):t}$ as input. In the special case $t=1$, the networks receive no input and reduce to two learnable parameters.
}. 
The base distribution $p(\y_{1:T})$ is typically fixed to be a simple factorized distribution such as an isotropic Gaussian, and is related to the distribution $p(\x_{1:T})$ through the standard change-of-variable formula between probability densities.
While the forward MAF transform ($\y_{1:T} \to \x_{1:T}$) converts a sequence of standard normal noise variables into a data sequence, the inverse transform ($\x_{1:T} \to \y_{1:T}$) removes temporal correlations and ``normalizes'' the data sequence. 

Although originally proposed for modeling static data (e.g., still images interpreted as a sequence of pixels), MAF can be applied along the temporal dimension of sequential data and is shown to improve video modeling performance \cite{marino2020improving}. This motivates us to consider the potential of MAF for sequential data compression.

\C{\subsection{On the Relation Between MAF and Sequence Compression}\label{sec:relation-btw-maf-and-compression}
In this section, we identify the commonality and difference between sequence modeling with MAF and sequence compression with predictive coding. On the one hand, MAF implements a core idea of 
decorrelation in transform coding, and the autoregressive transform underlying MAF resembles and generalizes that of traditional predictive coding. On the other hand, MAF does not consider the quantization and reconstruction error of lossy compression and is, therefore, not directly suitable for compression. 
Motivated by these two aspects, we will later on consider the model family of Variational Autoencoders (VAEs) \cite{kingma2013auto} that is more suited for compression (Sec.~\ref{sec:sequence-ntc}) but reintroduce the MAF-style autoregressive transform into the encoding and decoding procedure of the VAE (Sec.~\ref{sec:hybrid-method-stat}), resulting in a hybrid model that forms the basis of our proposal.

We begin by drawing conceptual connections between MAF and transform coding, the predominant paradigm of lossy compression. 
In transform coding, the data is first transformed to the domain of transform coefficients via a function $G: \x \to \y$, and the resulting coefficients $\y$ are compressed after scalar quantization. Although this two-stage approach is theoretically suboptimal compared to vector quantization \cite{gersho2012vector}, it has considerably lower computational complexity and is the default approach used in modern media compression algorithms \cite{goyal2001theoretical}. Conventionally, the transform $G$ is chosen to be orthogonal, and the rate-distortion-optimal transform is often characterized by its ability to \textit{decorrelate} the transform coefficients, i.e., $\operatorname{cov}(\y_i, \y_j) = 0$, for $i\neq j$ \cite{goyal2001theoretical}.
The idea of decorrelation is also a guiding principle behind data modeling with MAF or normalizing flow in general. 
}
Specifically, training a normalizing flow is, in fact, equivalent to decorrelating and ``normalizing'' the data distribution into a simple base distribution. 
Consider a flow model $p(\x)$ in the data space induced by passing a noise base distribution $p(\y)$ through a (forward) flow transform $F: \y \to \x$. 
If $p_\text{data}(\x)$ is the true data-generating distribution, then
the following relation holds \cite{papamakarios2017masked},
\begin{align}
    \text{KL}[ p_\text{data}(\x) \| p(\x) ] = \text{KL}[  F^{-1}(p_\text{data}(\x)) \| p(\y)], \label{eq:flow-duality}
\end{align}
where $F^{-1}(p_\text{data}(\x)) $ denotes the distribution that $p_\text{data}(\x)$ would follow when passed through the inverse transform $F^{-1}$.
In other words, training a flow model by maximum-likelihood is equivalent to fitting the ``normalized'' data distribution''  $F^{-1}(p_\text{data}(\x))$ to the base distribution $p(\y)$. \C{The connection to transform coding is then clear: the normalizing transform $F^{-1}$ plays a similar role to the transform $G$, $F^{-1}(p_\text{data}(\x))$ is the empirical distribution of ``transform coefficients'' $\y$, and $p(\y)$ corresponds to the factorized entropy model in transform coding.}

\C{Moreover, the autoregressive transform used by MAF is closely related to that of predictive coding and generalizes the latter. 
We can view predictive coding as implementing an autoregressive transform between the residual sequence and data sequence:
\begin{align}
    \y_t = \x_t - h_\mu(\hatx_{t-1}); \quad \texttt{// encode}; \label{eq:predictive-encoding-as-MAF} \\
    \hatx_t = h_\mu(\hatx_{t-1}) + \haty_t; \quad \texttt{// decode} , 
    \label{eq:predictive-decoding-as-MAF}
\end{align}
for $t=2,...,T$ \footnote{At time $t=1$, $\x_1$ is compressed and reconstructed as $\hatx_1$ separately without reference to any context frame.}.
In the language of predictive coding from Sec.~\ref{sec:background}, $h_\mu(\hatx_{t-1})=\barx_t$ is the prediction of the current frame, and $\y_t$ is the residual $\br_t$. 
The resulting transform can be seen as a special case of the more general shift-and-scale transform used by MAF in Eq.~\ref{eq:MAF}, if we fix $h_\sigma \equiv 1$ and limit $\hatx_{<t}$ to only the most recent (reconstructed) frame.

We now discuss the difference between sequence modeling and compression and the reasons why MAF may not be directly suitable for lossy compression.
In our discussion above, we have left out the issue of quantization. Unlike in sequence modeling, in lossy compression, the noise variable $\y_t$ must be communicated to the decoder in a lossy manner, e.g., via quantization $\haty_t = Q(\y_t)$ followed by (lossless) entropy coding. Similarly, we no longer have access to the history of ground truth frames $\x_{<t}$, but only their lossy reconstructions $\hatx_{<t}$.
Taking this into account, it is possible to construct a transform coding procedure using a learned MAF transform: 1. apply the learned $(h_\mu, h_\sigma)$ networks to compute the noise $\y_t$ via
\begin{align}
    \y_t = \frac{\x_t - h_\mu(\hatx_{t-1})}{h_\sigma(\hatx_{t-1})},\label{eq:predictive-encoding-as-MAF-generalized}
\end{align}
similar to Eq.~\ref{eq:predictive-encoding-as-MAF}; 2. quantize the noise to $\haty_t = Q(\y_t)$; 3.  compute the reconstruction $\hatx_t$ via
\begin{align}
    \hatx_t = h_\mu(\hatx_{t-1}) + h_\sigma(\hatx_{t-1}) \odot \haty_t,  \label{eq:predictive-decoding-as-MAF-generalized}
\end{align}
similar to Eq.~\ref{eq:predictive-decoding-as-MAF}, and finally, 4. iterate the above over time steps $t=1,2,...,T$.
}
Unfortunately, this simple transform coding procedure comes with a few practical drawbacks and is generally suboptimal in terms of rate-distortion performance.
First, since the flow transform $F$ is only trained to maximize the data likelihood, the resulting reconstruction error $\mathcal{D}(\x_{1:T}, \hatx_{1:T})$ due to quantization is uncontrolled. Empirically, trained flow transforms are often close to singular and suffer numerical stability issues \cite{behrmann2020understanding}, resulting in large reconstruction error $\mathcal{D}(\x, F((F^{-1}(\x))))$ even without the quantization step (despite $F$ being invertible in theory).
Secondly, the invertibility of the transform $F$ places restrictions on the kinds of computation allowed. This often necessitates deeper and more expensive neural network transforms to achieve similar expressivity to unconstrained neural network transforms. State-of-the-art normalizing flow models such as GLOW \cite{kingma2018glow} often require a deep stack of bijective transforms and are computationally much more expensive than comparable VAEs or GANs, making them potentially less suited for real-time video transmission applications.

\C{Lastly, we do note, however, that a connection exists between density modeling and the \emph{lossless} compression of quantized data through the technique of dequantization \cite{uria2013rnade}, and the latter has been used extensively in training and evaluating normalizing flows. Moreover, under fine quantization, the negative log density under a normalizing flow model can be recovered as the NELBO of a particular latent variable model, 
allowing bits-back coding to be applied for lossless compression \cite{ho2019compression}.
}

\subsection{Latent Variable Models for Learned Sequence Compression}\label{sec:sequence-ntc}

\C{We can overcome the suboptimality of normalizing flows for lossy compression by switching to another class of generative models. 
In this section, we motivate sequence compression with latent variable models, particularly VAEs, that can be trained to perform nonlinear transform coding \cite{balle2020nonlinear} and optimize for rate-distortion performance in an end-to-end manner. } From this generative modeling perspective, we give a detailed account of the probabilistic structure of the sequential latent variable model for learned online video compression, in particular, the correspondence between the data compression process and inference-generative process.

To motivate this approach, consider transform coding with a pair of flexible (but not necessarily invertible) transforms $f$  (``encoder'') and $g$ (``decoder'') that map between the video $\x_{1:T}$ and its transformed representation $\barz_{1:T} = f(\x_{1:T})$. Given sufficiently expressive $f$ and $g$,  quantization can be performed by element-wise rounding to the nearest integer, $\round{\cdot}$, resulting in the following R-D objective,
\begin{align}
\mathcal{L} = \mathbb{E}_{\x_{1:T} \sim p_\text{data}}[ \mathcal{D}(\x_{1:T}, g(\round{f(\x_{1:T})})) + \beta \mathcal{R}(\round{ f(\x_{1:T}) }) ].
\end{align}
Following \cite{balle2017end},
we approximate the rounding operations
by uniform noise injection to enable gradient-based optimization during training.
The resulting, relaxed version of the above R-D objective can be shown to be equal to the expected Negative Evidence Lower BOund (NELBO) of a particular \emph{compressive VAE} model, described below \cite{balle2017end,theis2017lossy}:
\begin{multline}
    \tilde{\mathcal{L}} = \mathbb{E}_{\x_{1:T} \sim p_\text{data}}[ \mathbb{E}_{q(\z_{1:T}|\x_{1:T})} [ -\log p(\x_{1:T} | \z_{1:T} )
    \\
     -\log p(\z_{1:T}) ] ]. \label{eq:overall-nelbo}
\end{multline}
In this compressive VAE model, the 
\C{noisy quantization}, 
$\round{\barz_{1:T}} \approx \barz_{1:T} + \mathbf{u}_{1:T},  \mathbf{u}_{1:T} \sim \mathcal{U}(-
\textbf{0.5}, \textbf{0.5})$, 
\C{is equivalently obtained by sampling from a}
particular variational posterior distribution $q(\z_{1:T}|\x_{1:T}) = \mathcal{U}(\barz_{1:T}-0.5, \barz_{1:T}+0.5)$, i.e., a unit-width uniform distribution whose mean $\barz_{1:T}$ is predicted by an amortized inference network $f$.
The likelihood $p(\x_{1:T}|\z_{1:T})$ follows a Gaussian distribution with fixed diagonal covariance  $\tfrac{\beta}{2 \log 2}  \mathbf{I}$, and mean $\hatx_{1:T} = g(\z_{1:T})$ computed by the generative network $g$, so that the negative log-likelihood $-\log p(\x_{1:T} | \z_{1:T} )$ equals the squared error distortion $\tfrac{1}{\beta} \|\x_{1:T} -\hatx_{1:T} \|^2$
weighted by $\tfrac{1}{\beta}$.
Following \cite{balle2017end}, the prior density $p(\z_{1:T})$ is parameterized to interpolate its discretized version $P(\z_{1:T})$, so that given integer valued $\z_{1:T}$, the negative log density $-\log p(\z_{1:T})$ measures the code length $-\log P(\z_{1:T})$ assigned by the entropy model $P ({\z_{1:T}})$. By minimizing the approximate R-D objective of Eq.~\ref{eq:overall-nelbo} with respect to the parameters of $f$, $g$, and $p( \z_{1:T})$, training an end-to-end neural compression model is thus equivalent to learning a VAE by maximum-likelihood and amortized variational inference. 
Note this is in contrast to the maximum-likelihood estimation of a normalizing flow model, which does not account for the distortion of lossy compression and results in suboptimal rate-distortion performance (as discussed in Sec.~\ref{sec:relation-btw-maf-and-compression}).

Given a compressive VAE, the compression of a data sequence $\x_{1:T}$ via transform coding closely corresponds to an inference-generation pass through the VAE, described in the following steps.
\textbf{1. encoding:} The encoder passes $\x_{1:T}$ through $f$ to obtain a transformed representation $\barz_{1:T}=f(\x_{1:T})$, thus computing the mean parameters $\barz_{1:T}$ of the variational distribution $q(\x_{1:T}|\z_{1:T})$ by amortized variational inference; \textbf{2. quantization and entropy coding:} A posterior sample $\z_{1:T}$ is drawn, i.e., $\z_{1:T} \sim q(\z_{1:T}|\x_{1:T})$ (or, deterministically computed as $\z_{1:T} = \round{\barz_{1:T}}$ at test time), and a bit-string encoding $\boldsymbol{\xi}$ of $\z_{1:T}$ is transmitted under the entropy model $P({\z_{1:T}})$; \textbf{3. decoding:} The decoder decodes $\z_{1:T}$ from bit-string $\boldsymbol{\xi}$ using the entropy model $P({\z_{1:T}})$, then computes the reconstruction by $\hatx_{1:T}=g(\z_{1:T})$, corresponding to the mean parameters of the Gaussian likelihood model $p( \x_{1:T}| \z_{1:T})$. Note that step 3 (decoding) is analogous to sampling from the generative model, but without adding diagonal Gaussian noise to $\hatx_{1:T}$ dictated by the Gaussian likelihood $p( \x_{1:T}| \z_{1:T})$. Indeed, if the bitstring $\boldsymbol{\xi}$ consists of a sequence of purely random bits, then it is well known that decoding $\boldsymbol{\xi}$ under the entropy model $P({\z_{1:T}})$ produces a sample  $\z_{1:T} \sim P({\z_{1:T}})$ \cite{cover1999elements}. For this reason, in the following discussions, we occasionally blur the distinction between the (random) latent variable $\z_{1:T} \sim P({\z_{1:T}})$ and the quantized latent representation $\round{\barz_{1:T}}$ (as would be decoded from a bitstring $\boldsymbol{\xi}$) to simplify notation. 

A data compression process with such a learned transform coding algorithm also implicitly defines a VAE model for the data. Specifically, let us consider the compression procedure of an online video compression codec, in which individual frames $\x_t$ are transmitted sequentially. 
The encoding-decoding process is specified recursively as follows.
Given the ground truth current frame $\x_t$ and the previously reconstructed frames $\hatx_{<t}$, the encoder is restricted to be of the form $\barz_t = f(\x_t,\hatx_{<t})$, and similarly the decoder computes the reconstruction sequentially based on previous reconstructions and the current encoding, $\hatx_t = g(\hatx_{<t}, \round{\barz_t}))$. Existing codecs usually condition on a \emph{single} reconstructed frame, substituting $\hatx_{<t}$ by $\hatx_{t-1}$ in favor of efficiency.
In the language of generative modeling and variational inference, the sequential encoder corresponds to a variational posterior of the form $q(\z_t|\x_t, \z_{<t})$, i.e., filtering,
and the sequential decoder corresponds to the likelihood $p(\x_t|\z_{\leq t})=\mathcal{N}(\hatx_t, \tfrac{\beta}{2\log 2}  \mathbf{I})$;  in both distributions, the probabilistic conditioning on $\z_{<t}$ is based on the observation that $\hatx_{t-1}$ is a deterministic function of $\z_{<t}$, if we identify $\round{ \barz_{t} }$ with the random variable $\z_{t}$ and unroll the recurrence $\hatx_t = g(\hatx_{<t}, \z_{t})$.  As we show, all sequential compression approaches considered in this work follow this paradigm and implicitly define generative models of the data as $ p(\x_{1:T}) \propto  p(\x_{1:T} | \z_{1:T}) p(\z_{1:T}) = p(\z_{1:T}) \prod_t p(\x_t | \z_{\leq t})$, where the likelihood model $p(\x_t | \z_{\leq t})$ at each time step is centered on the reconstruction $\hatx_t$. The key difference is in the definition of the decoding transform for computing $\hatx_t$ as a (stochastic) function of $\hatx_{<t}$ and $\z_t$.

\C{\subsection{Hybrid Model with Stochastic Temporal Autoregressive Transform}\label{sec:hybrid-method-stat}
Having laid out the general approach for end-to-end learned video compression with sequential VAEs, we specify the per-frame likelihood model $p(\x_t| \z_{\leq t})$ by revisiting the autoregressive transform of MAF from Sec.~\ref{sec:relation-btw-maf-and-compression}. }
The resulting model 
captures several existing low-latency neural compression methods as specific instances \cite{lu2019dvc, liu2019learned, lin2020m, lu2020content, agustsson2020scale,yang2020autoregressive} and gives rise to the exploration of new models.
Consider computing the reconstruction $\hatx_t$ using the forward temporal autoregressive transform of a MAF as in Eq.~\ref{eq:predictive-decoding-as-MAF-generalized},
\begin{align}
    \hatx_t = h_\mu(\hatx_{< t}) + h_\sigma(\hatx_{< t}) \odot \haty_t.
    \label{eq:MAF-decoding-transform}
\end{align}
As follows, we augment it with a latent variable $\z_t$. We may interpret the reconstructed $\haty_t$ as being decoded from the random variable $\z_t$ given some prior $p(\z_t)$, and a decoding transform $g_z$; additionally, $\z_t$ may enter into the computation of the shift $h_\mu$ and scale $h_\sigma$ transforms. By combining a sequential latent variable model with temporal autoregressive transforms, we therefore arrive at the most general form of the proposed \emph{stochastic temporal autoregressive transform}:

\begin{align}
    \hatx_t = h_\mu(\hatx_{< t}, \z_t) + h_\sigma(\hatx_{< t}, \z_t) \odot g_z(\z_t).
    \label{eq:master}
\end{align}
In this work, we only consider the common Mean Squared Error (MSE) distortion for simplicity. Therefore the above decoding transform computes the mean of a diagonal Gaussian frame observation model $p(\x_t | \z_{\leq t})$, with a fixed diagonal co-variance parameterized by the desired Lagrange multiplier $\beta$ (see explanation at the end of section \ref{sec:relation-btw-maf-and-compression}). We note, however, other kinds of distortion functions such as MS-SSIM are possible, resulting in a different form of $p(\x_t | \z_{\leq t})$ parameterized by $\hatx_t$ \cite{balle2017end}.

\begin{figure*}[t!]
    \centering
    \includegraphics[width=\linewidth]{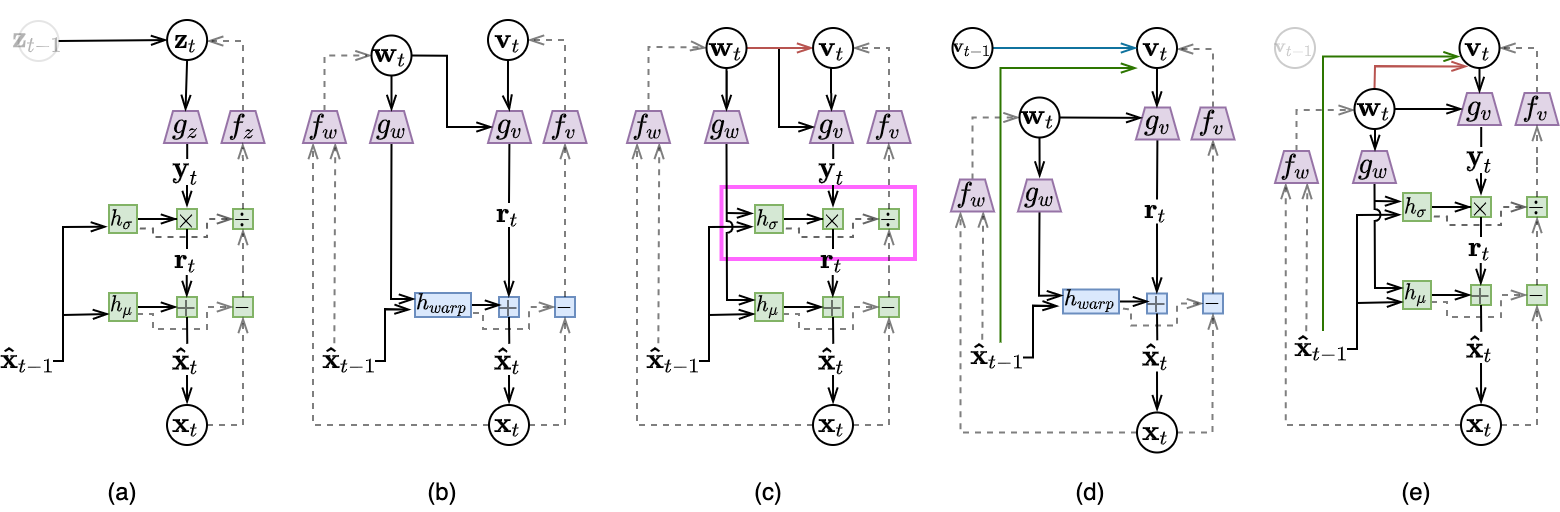}
    \caption{\textbf{Model diagrams} for the generative and inference procedures of the current frame $\x_t$, for various neural video compression methods. Random variables are shown in circles; all other quantities are deterministically computed; solid and dashed arrows describe computational dependency during generation (decoding) and inference (encoding), respectively. Purple nodes correspond to neural encoders (CNNs) and decoders (DCNNs), and green nodes implement temporal autoregressive transform.
    \textbf{(a)} TAT; \textbf{(b)} SSF; \textbf{(c)} STAT or STAT-SSF; the magenta box highlights the additional proposed scale transform absent in SSF, and the red arrow from $\w_t$ to $\bv_t$; highlights the proposed (optional) structured prior. \textbf{(d)} SSF-TP/SSF-TP+ and \textbf{(e)} STAT-SSF-SP-TP+ illustrate the temporal prior extension based on our proposal; the blue arrow shows the temporal dependency on the previous residual latent $\bv_{t-1}$, and the green arrow highlights the improved dependency on the previous reconstructed frame $\hatx_{t-1}$.
    }
    \label{fig:model-diagram}
\end{figure*}

This stochastic decoder model has several advantages over existing generative models for compression, such as simpler normalizing flows or sequential VAEs. First, the stochastic autoregressive transform $h_\mu(\hatx_{< t}, \z_t)$ involves a latent variable $\z_t$ and is therefore more expressive than a deterministic transform \cite{schmidt2018deep,schmidt2019autoregressive}.
Second, compared to MAF, which directly models $\y_t$, the additional nonlinear transform $g_z(\z_t)$ enables more expressive residual noise, reducing the burden on the prior by allowing it to model a simpler distribution of $\z_t$. Finally, as visualized in the video compression example in Figure~\ref{fig:qulitative-scale}, the scale parameter computed by $h_\sigma$ effectively acts as a gating mechanism, determining how much variance is explained in terms of the autoregressive transform and the residual noise distribution. This provides an added degree of flexibility, in a similar fashion to how RealNVP improves over NICE \cite{dinh2014nice, dinh2016density}.

Our approach is inspired by \cite{marino2020improving}, who analyzed a restricted version of the model in Eq.~\ref{eq:master}, aiming to hybridize autoregressive flows and sequential latent variable models for video modeling. In contrast to the stochastic transform in Eq.~\ref{eq:master}, 
the hybrid model in \cite{marino2020improving} is based on applying a \emph{deterministic} temporal autoregressive transform (as in MAF) to a sequence of residual noise variables $\y_{1:T}$ modeled by a sequential VAE, $p(\y_{1:T}) \propto p(\y_{1:T} | \z_{1:T}) p(\z_{1:T})$. The data distribution $p(\x_{1:T})$ under the resulting model (after applying the change-of-variable formula to the density of $p(\y_{1:T})$) does not admit a simple conditional likelihood distribution like $\x_t|\hatx_t \sim \mathcal{N}(\hatx_t, \tfrac{\beta}{2\log 2}  \mathbf{I})$, and maximum-likelihood training of $p(\x_{1:T})$ is not directly aligned with the R-D objective of video compression. We note that, however, the learned MAF transform of such a model may be used by a transform coding algorithm in a manner similar to our discussion in Sec.~\ref{sec:relation-btw-maf-and-compression}, but the resulting algorithm is likely to be suboptimal in R-D performance and faces similar issues with decoding.

\subsection{Example Models}\label{sec:models}

Next, we will show that the general framework expressed by Eq.~\ref{eq:master} captures a variety of state-of-the-art neural video compression schemes and gives rise to extensions and new models. 

\textbf{Temporal Autoregressive Transform (TAT).}
The first special case among the class of models that are captured by Eq.~\ref{eq:master} is the autoregressive neural video compression model by Yang et al. \cite{yang2020autoregressive}, which we refer to as temporal autoregressive transform (TAT). Shown in Figure \ref{fig:model-diagram}(a), the decoder $g$ implements a deterministic scale-shift autoregressive transform of decoded noise $\haty_t$,
\begin{align}
    \hatx_t = g(\hatx_{t-1}, \z_t) = h_\mu(\hatx_{t-1}) + h_\sigma(\hatx_{t-1}) \odot \haty_t, \quad \haty_t = g_z(\z_t).
    \label{eq:decoder-tat}
\end{align}
The encoder $f$ inverts the transform to decorrelate the input frame $\x_t$ into $\bary_t$ and encodes the result as $\barz_t = f(\x_t,\hatx_{t-1}) = f_z(\bary_t)$, where $\bary_t =\frac{\x_t -h_\mu(\hatx_{t-1})}{h_\sigma(\hatx_{t-1})}$. The shift $h_\mu$ and scale $h_\sigma$ transforms are parameterized by neural networks, $f_z$ is a convolutional neural network (CNN), and $g_z$ is a deconvolutional neural network (DCNN) that approximately inverts $f_z$.

The TAT decoder is a simple version of the more general stochastic autoregressive transform in Eq~\ref{eq:master}, where $h_\mu$ and $h_\sigma$ lack latent variables.
Indeed, focusing on the generative process of $\hatx$, TAT implements the model proposed by \cite{marino2020improving}, transforming $\y$ into $\hatx$ by a MAF. However, the generative process underlying lossy compression (see Section~\ref{sec:sequence-ntc}) adds additional white noise to $\hatx$, with $\x:=\hatx + \boldsymbol{\epsilon}, \boldsymbol{\epsilon}\sim \mathcal{N}(\mathbf{0}, \tfrac{\beta}{2\log 2} \mathbf{I})$. Thus, the generative process from $\y$ to $\x$ is no longer invertible nor corresponds to an autoregressive flow.
Nonetheless, TAT was shown to better capture the low-level dynamics of video frames than the autoencoder $(f_z, g_z)$ alone, and the inverse transform decorrelates raw video frames to simplify the input to the encoder $f_z$ \cite{yang2020autoregressive}.

\textbf{DVC \cite{lu2019dvc}, Scale-Space Flow (SSF, \cite{agustsson2020scale}), among others \cite{liu2019learned, lin2020m, lu2020content}.}
The second class of models captured by Eq.~\ref{eq:master} belong to the neural video compression framework based on predictive coding; both models make use of two sets of latent variables $\z_{1:T} = \{ \w_{1:T}, \bv_{1:T}  \}$ to capture different aspects of information being compressed,
where $\w$ captures estimated motion information used in the warping prediction, and $\bv$ helps capture residual error not predicted by warping the previous reconstruction frame.  

Like most classical approaches to video compression by predictive coding, the reconstruction transform in the above models has the form of a prediction shifted by residual error (decoded noise), and lacks the scaling factor $h_\sigma$ compared to the autoregressive transform in Eq.~\ref{eq:master}
\begin{align}
    \hatx_t = h_{warp}(\hatx_{t-1}, g_w(\w_t)) + g_v(\bv_t, \w_t). \label{eq:encoder-ssf}
\end{align}
Above, $g_w$ and $g_v$ are DCNNs, $\mathbf{o}_t := g_w(\w_t)$ has the interpretation of an estimated optical flow (motion) field, $h_{warp}$ denotes warping \cite{glasbey1998review}, the $h_\mu$ of Eq.~\ref{eq:master} is defined by the composition $h_\mu(\hatx_{t-1}, \w_t) := h_{warp}(\hatx_{t-1}, g_w(\w_t))$,
and the residual $\br_t := g_v(\bv_t, \w_t) = \hatx_t - h_{warp}(\hatx_{t-1}, \mathbf{o}_t)$ represents the prediction error unaccounted for by warping.
DVC \cite{lu2019dvc} only makes use of $\bv_t$ in the residual decoder $g_v$, and performs simple 2D warping by bi-linear interpretation;  Lin et al. \cite{lin2020m} make use of multiple reference frames $\hatx_{(t-3):t}$ for estimating the optimal flow (motion) field; 
SSF \cite{agustsson2020scale} augments the optical flow (motion) field $\mathbf{o}_t$ with an additional scale field,
and applies scale-space-warping to the progressively blurred versions of $\hatx_{t-1}$ to allow for uncertainty in the warping prediction.
The encoding procedure in the above models computes the variational mean parameters as $\barw_t = f_w(\x_t, \hatx_{t-1}), \barv_t = f_v(\x_t - h_{warp}(\hatx_{t-1}, g_w(\w_t)))$, corresponding to a structured posterior $q(\z_t | \x_t, \z_{<t}) = q(\w_t| \x_t, \z_{<t}) q(\bv_t| \w_t, \x_t, \z_{<t})$. 
We illustrate the above generative and inference procedures in Figure~\ref{fig:model-diagram}(b).

\subsection{Proposed Models (Base Versions)}\label{sec:base-stat-models}
\begin{figure*}[t!]
    \centering
    \includegraphics[width=1\linewidth]{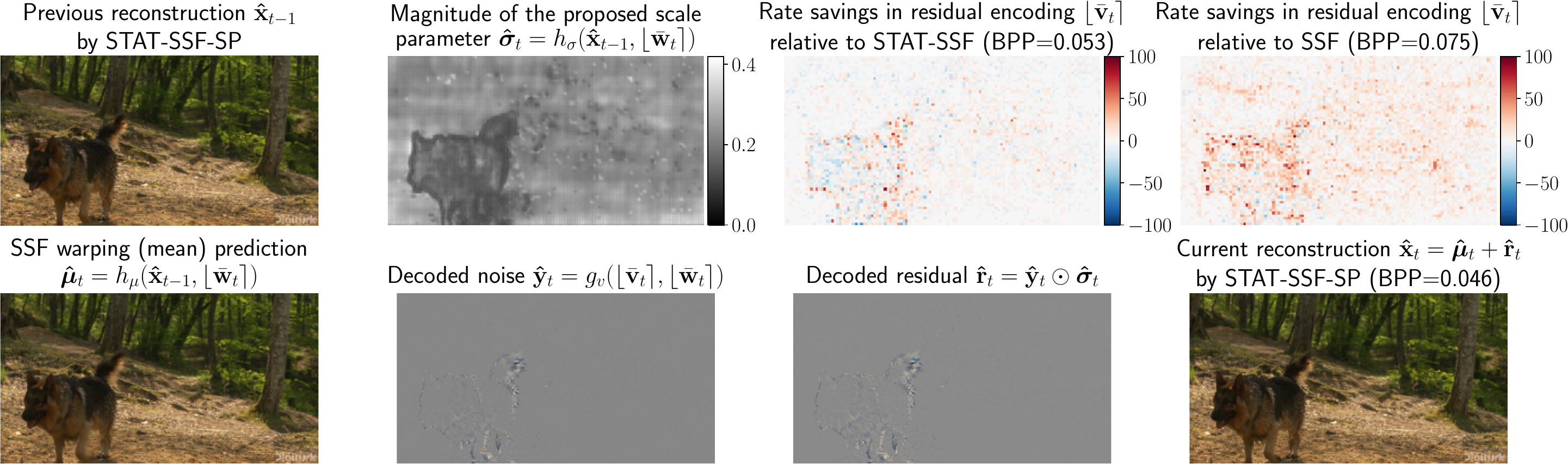}
    \caption{\C{\textbf{Visualizing the encoding/decoding computation of the STAT-SSF-SP model} on one frame of UVG video ``Shake-NDry''. See Fig.~\ref{fig:model-diagram}(c) for the model's computation diagram.
    In this example, the warping prediction $\bm{\hat\mu}_t$ (bottom, first) incurs a large error around the dog's moving contour but models the mostly static background well, with the residual latents $\round{\barv_t}$ taking up an order of magnitude higher bit-rate than $\round{\barw_t}$. %
    The proposed scale parameter $\bm{\hat\sigma}_t$ (top, second) gives the model extra flexibility when combining the noise $\haty_t$ 
    (bottom, second)
    with the warping prediction $\bm{\hat\mu}_t$ 
    to form the reconstruction $\hatx_t = \bm{\hat\mu}_t + \bm{\hat\sigma}_t \odot \haty_t$ (bottom, fourth). The scale $\bm{\hat\sigma}_t$ downweights contribution from the noise $\haty_t$ in the foreground where it is very costly, and reduces the residual bit-rate $\mathcal{R}(\round{\barv_t})$ (and thus the overall bit-rate) compared to STAT-SSF and SSF, as illustrated in the third and fourth figures in the top row.
    The (BPP, PSNR) performance for STAT-SSF-SP, STAT-SSF, and SSF \cite{agustsson2020scale} are (0.046, 36.97), (0.053, 36.94), and (0.075, 36.97), respectively. Thus, STAT-SSF and SSF here have comparable reconstruction quality to STAT-SSF-SP but worse bit-rate.
    }
    }
    \label{fig:qulitative-scale}
\end{figure*}

Finally, we consider a version of stochastic temporal autoregressive transform (Eq.~\ref{eq:master}) in the context of predictive video coding,
\begin{align}
    \hatx_t = h_\mu(\hatx_{t-1}, \w_t) + h_\sigma(\hatx_{t-1}, \w_t) \odot g_v(\bv_t, \w_t).
    \label{eq:master-video}
\end{align}
As in DVC and SSF, the latent variable $\z_t$ consists of two components $\bv_t, \w_t$, and the shift and scale parameters are computed using only the previous reconstruction $\hatx_{t-1}$. See Figure~\ref{fig:model-diagram}(c) for a diagram of the resulting generative model.
We study two main variants, categorized by how they implement $h_\mu$ and $h_\sigma$:
\begin{itemize}
    \item \textbf{STAT} uses DCNNs for $h_\mu$ and $h_\sigma$ as in Yang et al. \cite{yang2020autoregressive}, \C{but both networks receive the latent variable $\w_t$ as additional input, which helps guide the transform.
    In theory, the universal approximation property of neural networks should allow us to learn whichever flexible functions $(h_\mu, h_\sigma)$ achieve the best compression performance. However, in practice, we find the following variant based on warping to be more performant and parameter-efficient.}
    \item \textbf{STAT-SSF} is a %
    more domain-knowledge-driven
    variant of the above that still uses scale-space warping \cite{agustsson2020scale} in the shift transform, i.e., $h_\mu(\hatx_{t-1}, \w_t) = h_{warp}(\hatx_{t-1}, g_w(\w_t))$. This can also be seen as an extended version of the SSF model, whose shift transform $h_\mu$ is preceded by a new learned scale transform $h_\sigma$. \C{We motivated the scaling transform in Sec.~\ref{sec:hybrid-method-stat}, and provide a visualization of its effect in Fig.~\ref{fig:qulitative-scale}.}
\end{itemize}

\textbf{Structured Prior (SP). }
Besides improving the autoregressive transform (affecting the likelihood model for $\x_t$), one variant of our approach also improves the topmost generative hierarchy in the form of a more expressive latent prior $p(\z_{1:T})$, affecting the entropy model for compression. We observe that motion information encoded in $\w_t$ can often be informative of the residual error encoded in $\bv_t$. In other words, large residual errors $\bv_t$ incurred by the mean prediction $h_\mu(\hatx_{t-1}, \w_t)$ (e.g., the result of warping the previous frame $h_\mu(\hatx_{t-1}$)) are often spatially collocated with (unpredictable) motion as encoded by $\w_t$.

The original SSF model's prior factorizes as $p(\w_t, \bv_t) = p(\w_t) p(\bv_t)$ and does not capture such correlation. 
We, therefore, propose a structured prior by introducing conditional dependence between $\w_{t}$ and $\bv_t$, so that $ p(\w_t, \bv_t) =  p(\w_t) p(\bv_t | \w_t)$. At a high level, this can be implemented by introducing a new neural network that maps $\w_t$ to parameters of a parametric distribution of $p(\bv_t | \w_t)$ (e.g., mean and variance of a diagonal Gaussian distribution).
This results in variants of the above models, \textbf{STAT-SP} and \textbf{STAT-SSF-SP}, where the structured prior is applied on top of the proposed \textbf{STAT} and \textbf{STAT-SSF} models, respectively.

\subsection{Temporal Prior (TP) Extensions}

In previous sections and other similar works in variational compression\cite{agustsson2020scale,lu2019dvc,habibian2019video}, the prior model $p(\z_{1:T})$ typically assumes that the latent variables $\z_{1:T}$ are temporally factorized: $ p(\z_{1:T}) = \prod_{t=1}^T p(\z_t)$. However, such assumptions may be unrealistic for real world data, such as modeling natural video \cite{denton2018stochastic, li2018disentangled, marino2020improving}, where temporal dependencies may persist even after removing low-level motion information. To this end, we model these dependencies by introducing a temporal prior: $ p(\z_{1:T}) = p(\z_1) \prod_{t=2}^T p(\z_t|\z_{<t})$ to more accurately predict the density of the latent variable. Given the framework described in the previous section, we implement the prior of the video compression model with the following two types of temporal conditioning. Meanwhile, the motion latent $\w_t$ is not discussed here, as it usually has a much lower bitrate than $\bv_t$.

\textbf{Temporal Residual Conditioning $p(\bv_t|\bv_{t-1})$ (TP).} Most video compression methods are based on the Markov assumption, e.g., an optical flow's vector field is only conditioned on the most recent frame. Empirically, we find that the information content of the compressed flow field is much smaller than that of the residual,  $-\log p(\bv_t) \gg -\log p(\w_t)$. Therefore, we only place a temporal prior on $\bv_t$ and keep  $p(\w_t)$ temporally factorized for simplicity. Temporal conditioning for $\bv$ is implemented by using an additional neural network that takes $\bv_{t-1}$ as input to a conditional Gaussian prior for $\bv_t$. However, this conditioning scheme may require special handling of the initial frames because the individual image compression model for the I-frame (the first frame) does not have a ``residual.'' This indicates that such temporal conditioning can only start from the 2$^\textrm{nd}$ frame, as $p(\bv_t|\bv_{t-1})$ is only available for $t\geq 3$. Instead, we use a separate factorized prior $p(\bv_2)$ for the 2$^\textrm{nd}$ frame.

\textbf{Conditioning on the Previous Frame $p(\bv_t|\hatx_{t-1})$ (TP+).} Learning a temporal prior by only conditioning on the latent variable $\bv$  is challenging in practice. First, $\bv$ is a noisy quantity during training and only carries information of the previous frame's \emph{residual}, lacking explicit information of the previous frame.
Furthermore, the latent representation can change throughout training because of the noise injection, potentially complicating the learning process. Instead, we explore an alternative scheme, \textbf{TP+}, in which $\bv_t$ is conditioned on the previous reconstruction, $\hatx_{t-1}$, which maintains a less noisy, more informative, and more consistent feature representation throughout training and simplifies the learning procedure. In this scenario, the model also no longer requires the extra prior for $p(\bv_2)$, as in \textbf{TP}; moreover, the resulting prior $p(\bv_t|\hatx_{t-1}) = p(\bv_t| \w_{<t}, \bv_{<t})$ (since $\hatx_{t-1}$ is a deterministic function of $\z_{<t}$) 
offers a strictly more expressive probabilistic model than \textbf{TP}, $p(\bv_t|\bv_{t-1})$.

\subsection{Variable Bitrate Extensions}

Classical compression methods such as HEVC or AVC typically use \emph{one} compression model to compress a video for 52 different quality levels. For example, in the FFMPEG library, this can be achieved by varying the value of the Constant Rate Factor (CRF). 
However, current end-to-end trained neural codecs largely focus on optimal rate-distortion performance at a fixed bitrate.
As discussed, various neural video compression models \cite{agustsson2020scale, habibian2019video, lu2019dvc, liu2019learned} minimize the NELBO objective, $\mathcal{L} = \mathcal{D} + \beta\mathcal{R}$, where $\beta$ controls the rate-distortion tradeoff, and a separate compression  model needs to be optimized for each setting of $\beta$.
The overhead of training and deploying multiple models (e.g, 52 sets of neural network parameters for 52 quality levels) potentially makes neural video compression impractical. Therefore, we consider a variable-rate compression setting, where we train a single model with multiple $\beta$ values and adapt it to different quality factors at deployment time.

The analogy between the compression models and deep latent variable models (see Sec.~\ref{sec:sequence-ntc}) inspires us to draw on conditional variational autoencoders \cite{NIPS2015_8d55a249} that model conditional distributions instead of unconditional ones. By conditioning the networks on the quality control factor $B$ and generating random $Q$-samples during training, we can learn a single set of encoder, decoder, and prior model that operates near-optimally at different bitrates. Specifically, we draw on a proposal by \cite{choi2019variable} for images and generalize it to the class of neural video codecs studied here. 

In more detail, we define a quality control factor $B$, represented by a one-hot vector, as the condition of the compression model where each $B$ has a unique corresponding $\beta$. Then we can derive the new conditional optimization objective from equation \ref{eq:overall-nelbo}:
\begin{align}
  \phantom{\tilde{\mathcal{L}_\text{VBR}}}
  &\begin{aligned}
    \mathllap{\tilde{\mathcal{L}_\text{VBR}}} &= \mathbb{E}_{\x_{1:T} \sim p_{\text{data}}}\mathbb{E}_{q(\z_{1:T}|\x_{1:T},B)}\\
      &\qquad [-\log p(\x_{1:T} | \z_{1:T}, B) -\beta\log p(\z_{1:T} | B)]
  \end{aligned}\\
  & \equiv \mathcal{D}(B) + \beta \mathcal{R}(B)
\end{align}
During training, a $(B, \beta)$ pair is selected randomly for each training sample. Following \cite{choi2019variable}, we also replace all convolutions with conditional convolution to incorporate the $B$ variable. We found that one can technically use simpler methods (like directly reshaping and concatenating $B$ to the network hidden features) to achieve the same result.

\section{Related Work}
\label{sec:related-work}
We divide related work into three categories: neural image compression, neural video compression, and sequential generative models. 

\textbf{Neural Image Compression.} Considerable progress has been made by applying neural networks to image compression; here we focus on the lossy (instead of lossless) compression setting more relevant to us. Early works \cite{Toderici_2017_CVPR, johnston2018improved} leveraged LSTMs to model spatial correlations of the pixels within an image. \cite{balle2016endtoend, theis2017lossy} were among the first to train an autoencoder-stlye model with rate-distortion loss for end-to-end lossy image compression, and proposed uniform noise injection \cite{balle2016endtoend} or the straight-through estimator \cite{bengio2013estimating} to approximately differentiate through quantization.  The end-to-end approach based on uniform noise injection was given a principled \emph{probabilistic} interpretation as implementing a deep latent variable model, a variational autoencoder (VAE) \cite{kingma2013auto}.
In subsequent work, \cite{balle2018variational} encoded images with a two-level VAE architecture involving a scale hyper-prior, which can be further improved by autoregressive structures \cite{minnen2018joint,minnen2020channel} or by optimization at encoding time \cite{yang2020improving}. \cite{yang2020variational} and \cite{flamich2019compression} demonstrated competitive image compression performance without a pre-defined quantization grid.  
Recently, \cite{helminger2020lossy} attempted to apply normalizing flow to lossy image compression; noting the incompatibility between density modeling with flow and the R-D objective of lossy compression (which we discuss in more detail in Sec.~\ref{sec:relation-btw-maf-and-compression}), they essentially trained an invertible autoencoder with a R-D loss instead, resulting in superior rate-distortion performance in the high bitrate regime but otherwise inferior performance compared to the VAE approach \cite{balle2017end, balle2018variational} without the invertibility requirement.

\textbf{Neural Video Compression.}
Compared to image compression, video compression is a significantly more challenging problem, as statistical redundancies exist not only within each video frame (exploited by intra-frame compression) but also along the temporal dimension. Early works by \cite{wu2018video,9009574} and \cite{han2018deep} performed video compression by predicting future frames using a recurrent neural network, whereas \cite{chen2019learning} and \cite{8305033} used convolutional architectures within a traditional block-based motion estimation approach. These early approaches did not outperform the traditional H.264 codec and barely surpassed the MPEG-2 codec. \cite{lu2019dvc} adopted a hybrid architecture that combined a pre-trained Flownet \cite{dosovitskiy2015flownet} and residual compression, which leads to an elaborate training scheme.  \cite{habibian2019video} and \cite{liu2019learned} combined 3D convolutions for dimensionality reduction with  expressive autoregressive priors for better entropy modeling at the expense of parallelism and runtime efficiency. Our method extends a low-latency model proposed by \cite{agustsson2020scale}, which allows for end-to-end training, efficient online encoding and decoding, and parallelism.

\textbf{Sequential Deep Generative Models.}
We drew inspiration from a body of work on sequential generative modeling. Early deep learning architectures for dynamics forecasting involved RNNs \cite{chung2015recurrent}. \cite{denton2018stochastic} and \cite{babaeizadeh2017stochastic} used VAE-based stochastic models in conjunction with LSTMs to model dynamics. \cite{li2018disentangled} introduced both local and global latent variables for learning disentangled representations in videos. Other video generation models used generative adversarial networks (GANs) \cite{vondrick2016generating, lee2018stochastic} or autoregressive models and normalizing flows \cite{rezende2015variational, dinh2014nice, dinh2016density, kingma2018glow, kingma2016improved,papamakarios2017masked}. Recently, \cite{marino2020improving} proposed to combine latent variable models with autoregressive flows for modeling dynamics at different levels of abstraction, which inspired our models and viewpoints.  

\section{Experiments}
\label{sec:experiments}

In this section, we demonstrate the effectiveness of our models compared with baseline neural video compression solutions \cite{agustsson2020scale, lu2019dvc} as well as a state-of-the-art classic video codec when evaluated on two publicly available benchmark datasets. The backbone modules and training scheme for our models largely follow \cite{agustsson2020scale}.

Regarding the implementation of scale-space volume \cite{agustsson2020scale}, we adopted a Gaussian pyramid approach with successive convolution and downsampling. Compared with naively applying separate Gaussian kernels with exponentially increasing kernel width (referred to as \emph{Gaussian-blur} approach), the Gaussian pyramid-based implementation uses a fixed small convolution kernel,  yields similar scale-space volume, and is much more efficient in terms of computation. The pseudo-code is presented in Algorithm~\ref{alg} 

\begin{algorithm}[h]
\SetAlgoLined
\KwResult{\textit{ssv}: Scale-space 3D volume}
 \textbf{Input:} \textit{image} input image; $\sigma_0$ base scale; $M$ scale depth\;
 ssv = [image]\;
 kernel = Gaussian\_Kernel($\sigma_0$)\;
 \For{i=0 to M-1}{
  image = Conv(image, kernel)\;
  \eIf{i == 0}{
   ssv.append(image)\;
   }{
   scaled = image\;
   \For{j=0 to i-1}{
   scaled = UpSample2x(scaled)\;
   }
   ssv.append(scaled)\;
  }
  image = DownSample2x(image)\;
 }
 \Return Concat(ssv)
 \caption{Gaussian pyramid for scale space volume}
 \label{alg}
\end{algorithm}

\subsection{Training and Evaluation}
\label{sec: te}

\textbf{Training}. We train on the Vimeo-90k \cite{xue2019video} dataset as in previous works \cite{lu2019dvc,yang2020Learning,liu2019learned}. 
\C{We follow largely the same procedure as SSF \cite{agustsson2020scale}, and give more details in the supplementary material.
Table \ref{tab:table-model} summarizes our proposed models and the baseline methods whose published results we compare with.
}

\vspace{1em}

\noindent\textbf{Evaluation}. 
We report compression performance on two widely used datasets: \textbf{UVG}\footnote{UVG dataset is provided by \url{http://ultravideo.fi/} and under CC-by-NC 3.0 license.} \cite{mercat2020uvg} and \textbf{MCL\_JCV} \cite{wang2016mcl}. Both consist of raw videos in YUV420 format. UVG contains seven 1080p videos at 120fps with smooth and mild motions or stable camera movements. MCL\_JCV contains thirty 1080p videos at 30fps, which are generally more diverse, with a higher degree of motion and scene cuts.

We report the bitrate (bits-per-pixel, or BPP for short) and the reconstruction quality, measured in PSNR on 8-bit RGB space (same as in training), averaged across all frames. 
PSNR is a more challenging metric than MS-SSIM \cite{wang2003multiscale} for learned codecs \cite{lu2019dvc, agustsson2020scale, habibian2019video, yang2020Learning, yang2020feedback, johnston2019computationally}.
Since most existing neural compression methods assume video input in 8-bit RGB444 format (24 bits per pixel), we follow this convention by converting test videos from YUV420 to RGB444 in our evaluations for meaningful comparison. It is worth noting that HEVC is mainly designed for YUV420 with sub-sampled chroma components, which means that it can exploit the fact that chroma components in test videos are subsampled and thus gains an unfair advantage in the comparison against neural codecs. Nevertheless, we report the performance of HEVC as a reference.

\begin{table*}[ht]
\centering
\caption{\textbf{\C{Overview of various compression methods considered and the contexts in which they appear.}}}
\begin{tabular}{|c|c|c|}
\hline
\textbf{Model Name}   & \textbf{Category}  & \textbf{Remark}                                                                              \\
\hline\hline
\textbf{STAT-SSF}       & Proposed         & Proposed autoregressive transform with scale space flow                       \\
\hline
\textbf{STAT-SSF-SP}    & Proposed   & Same as above (STAT-SSF) but with structured prior \\
\hline
\textbf{STAT-SSF-SP-TP+}    & Proposed  &  \makecell{Structured prior and improved temporal prior, $p(\bv_t|\hatx_{t-1}, \w_t)$}\\
\hline
\hline
SSF          & Baseline    & Agustsson et al.\cite{agustsson2020scale}                                 \\
\hline
DVC          & Baseline      & Lu et al.\cite{lu2019dvc}                                                                 \\
\hline
VCII    & Baseline        & Wu et al.\cite{wu2018video} (trained on the Kinectics dataset)                                            \\
\hline
DGVC         & Baseline   & Han et al.\cite{han2018deep} modified for low-latency compression setup                                                        \\
\hline
TAT          & Baseline    & Yang et al.\cite{yang2020autoregressive}                                                    \\
\hline
HEVC    & Baseline    & FFMPEG-HEVC with RGB 4:4:4 color space input                                            \\
\hline
HEVC(YUV)    & Baseline     & FFMPEG-HEVC with YUV 4:2:0 color space input                                         \\
\hline
STAT         & Ablation       & STAT-SSF without optical flow   \\
\hline
SSF-SP       & Ablation       & SSF with structured prior                                       \\
\hline
SSF-TP    & Ablation & SSF with temporal prior conditioned on $\bv_{t-1}$, $p(\bv_t|\bv_{t-1})$\\
\hline
SSF-TP+    & Ablation & SSF with temporal prior conditioned on $\hatx_{t-1}$,  $p(\bv_t|\hatx_{t-1})$\\
\hline
\end{tabular}
\label{tab:table-model}
\end{table*}

\begin{figure*}[h]
    \centering
    \begin{subfigure}[]{.48\linewidth}
        \includegraphics[width=\linewidth]{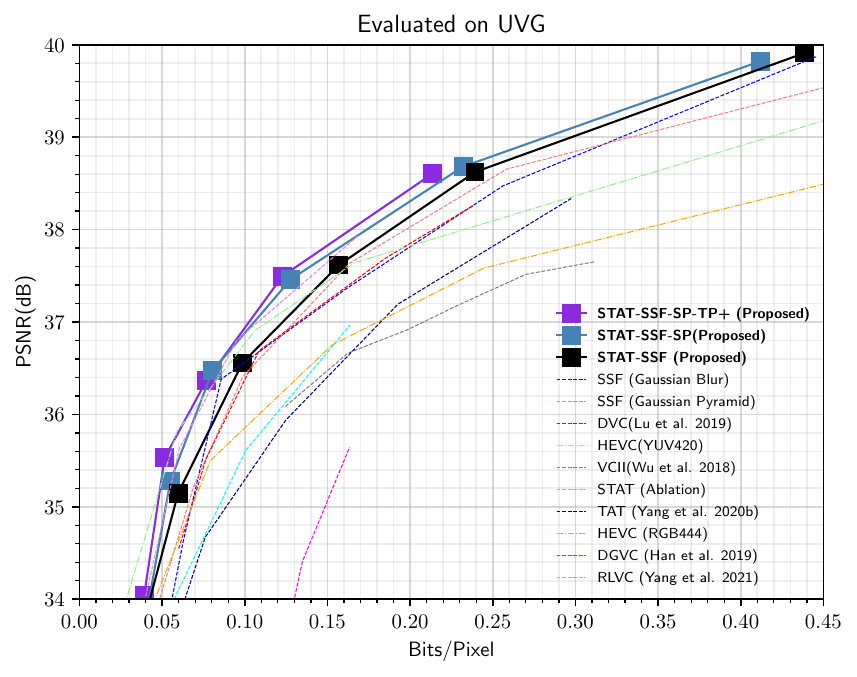}
        \caption{}
        \label{fig:uvg-rd}
    \end{subfigure}
    \begin{subfigure}[]{.48\linewidth}
        \includegraphics[width=\linewidth]{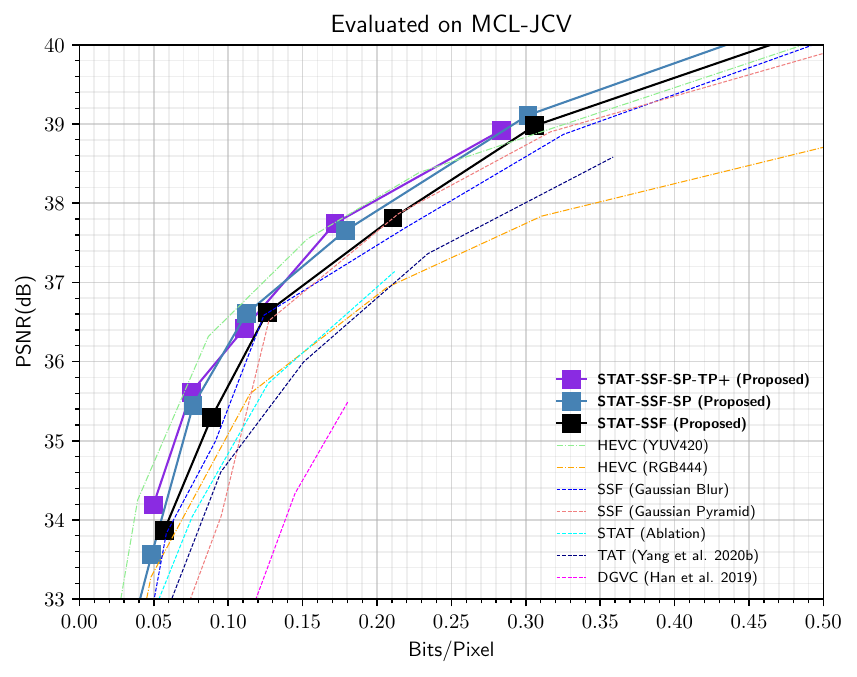}
        \caption{}
        \label{fig:mcl-rd}
    \end{subfigure}
    \caption{\textbf{Rate-Distortion Performance} of various models and ablations. Results are evaluated on \textbf{(a)} UVG and \textbf{(b)} MCL\_JCV datasets. All the learning-based models (except VCII \cite{wu2018video}) are trained on Vimeo-90k. STAT-SSF-SP-TP+ (proposed) achieves the best performance.}
    \vspace{-1em}
    \label{fig:main-rd}
\end{figure*}

\begin{figure*}[ht]
\centering
\begin{subfigure}[h]{.19\textwidth}
  \centering
  \includegraphics[width=\linewidth, trim={0 110px 0 0}, clip]{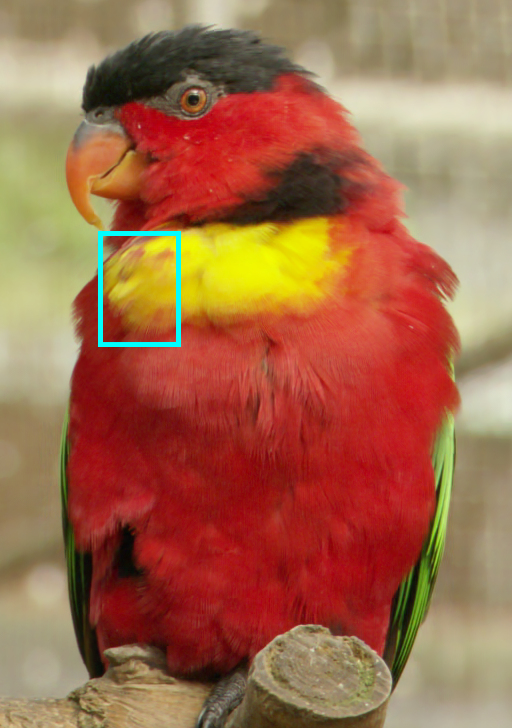} %
  \includegraphics[width=\linewidth, trim={0 12px 0 5px}, clip]{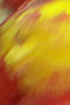}
  \caption{HEVC; \\ $\text{BPP}=0.087$, \\ $\text{PSNR}=38.10$}
\end{subfigure}\hfill
\begin{subfigure}[h]{.19\textwidth}
  \centering
  \includegraphics[width=\linewidth, trim={0 80px 0 0}, clip]{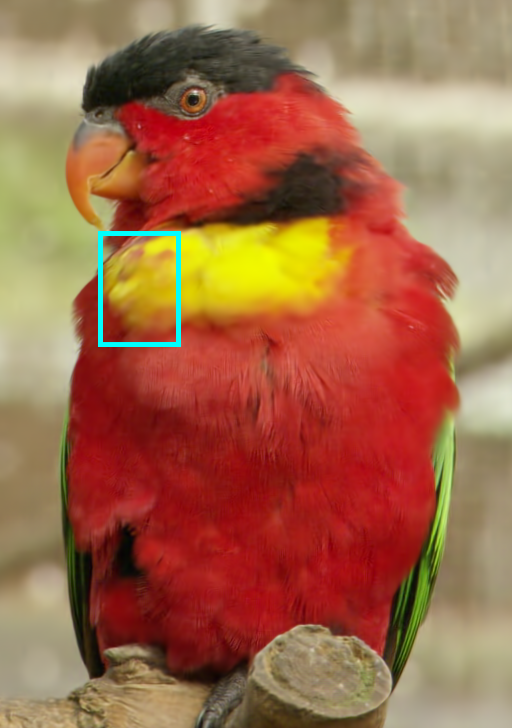}
  \includegraphics[width=\linewidth, trim={0 10px 0 2px}, clip]{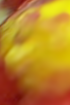}
  \caption{SSF; \\ $\text{BPP} = 0.082$, \\ $\text{PSNR}=37.44$}
\end{subfigure}\hfill
\begin{subfigure}[h]{.19\textwidth}
  \centering
  \includegraphics[width=\linewidth, trim={0 80px 0 0}, clip]{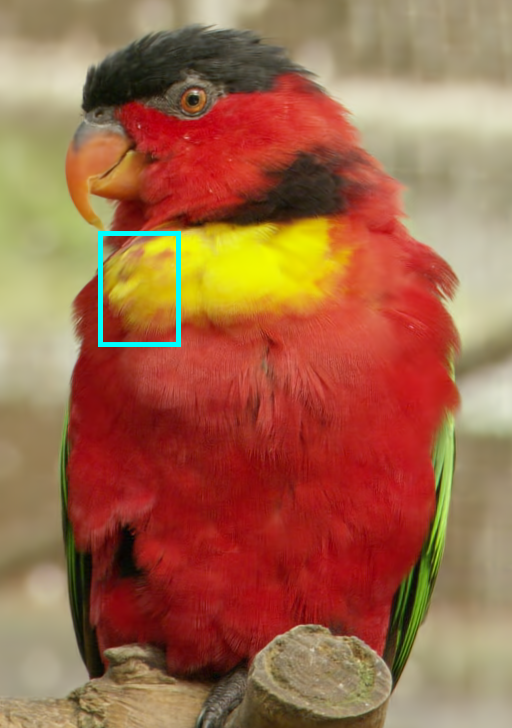}
  \includegraphics[width=\linewidth, trim={0 10px 0 2px}, clip]{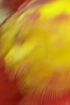}
  \caption{STAT-SSF (ours); \\ $\textbf{\text{BPP} = 0.077}$,\\ $\text{PSNR}=38.11$}
\end{subfigure}\hfill
\begin{subfigure}[h]{.19\textwidth}
  \centering
  \includegraphics[width=\linewidth, trim={0 80px 0 0}, clip]{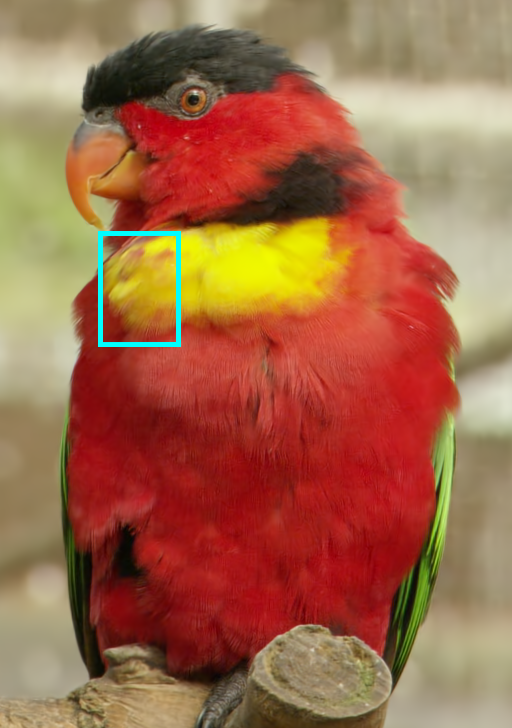}
  \includegraphics[width=\linewidth, trim={0 10px 0 2px}, clip]{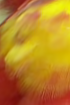}
  \caption{STAT-SSF-SP (ours); \\ $\textbf{\text{BPP} = 0.055}$,\\ $\text{PSNR}=38.10$}
\end{subfigure}\hfill
\begin{subfigure}[h]{.19\textwidth}
  \centering
  \includegraphics[width=\linewidth, trim={0 110px 0 0}, clip]{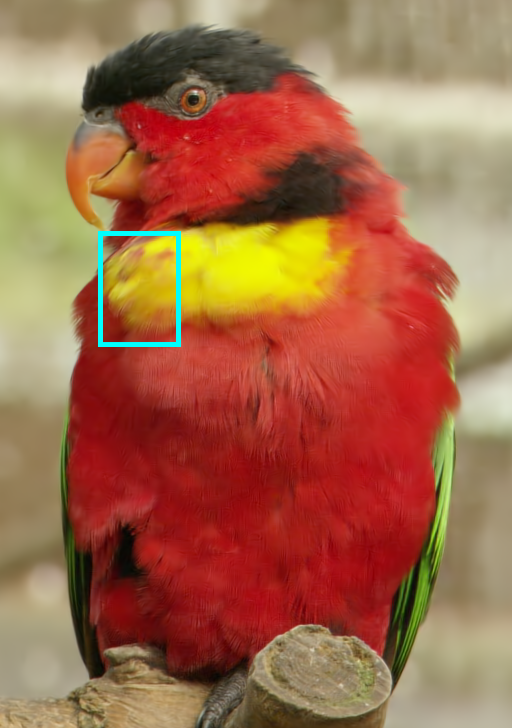}
  \includegraphics[width=\linewidth, trim={0 12px 0 5px}, clip]{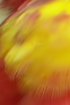}
  \caption{STAT-SSF-SP-TP+ \\ (ours); $\textbf{\text{BPP} = 0.054}$,\\ $\text{PSNR}=38.09$}
\end{subfigure}
\caption{\textbf{Qualitative comparisons} of various methods on a frame from MCL-JCV video 30. Figures in the bottom row focus on the same image patch on top. Here, models with the proposed scale transform (\textbf{STAT-SSF} and \textbf{STAT-SSF-SP}) outperform the ones without, yielding visually more detailed reconstructions at lower rates. The structured prior (\textbf{STAT-SSF-SP}) and temporal prior (\textbf{STAT-SSF-SP-TP+}) reduce the bitrate further. }
\vspace{-1em}
\label{fig:qualitative-reconstructions}
\end{figure*}

\begin{figure*}[h]
    \centering
    \begin{subfigure}[]{.48\linewidth}
        \includegraphics[width=\linewidth]{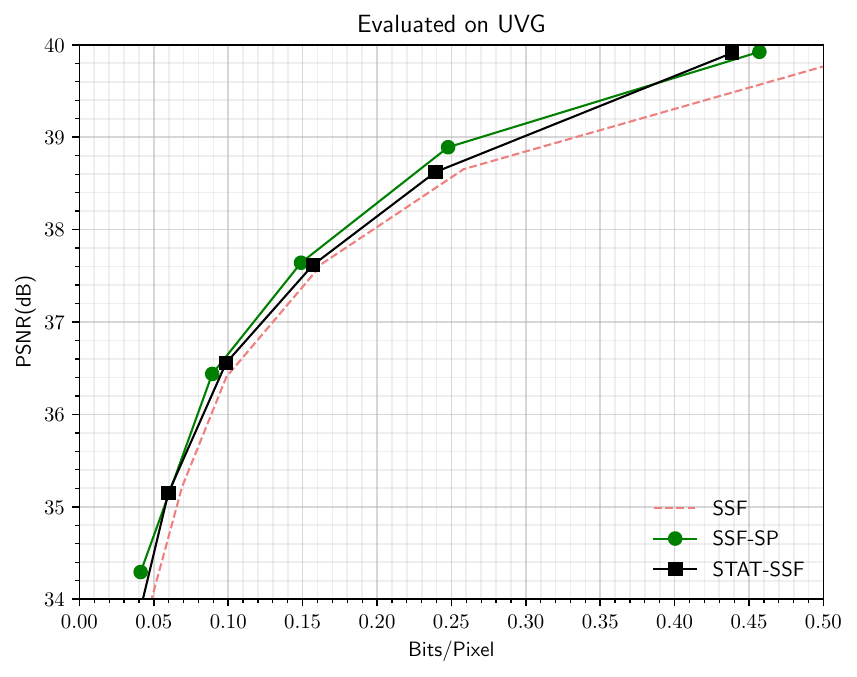}
        \caption{}
        \label{fig:ablation}
    \end{subfigure}
    \begin{subfigure}[]{.48\linewidth}
        \includegraphics[width=\linewidth]{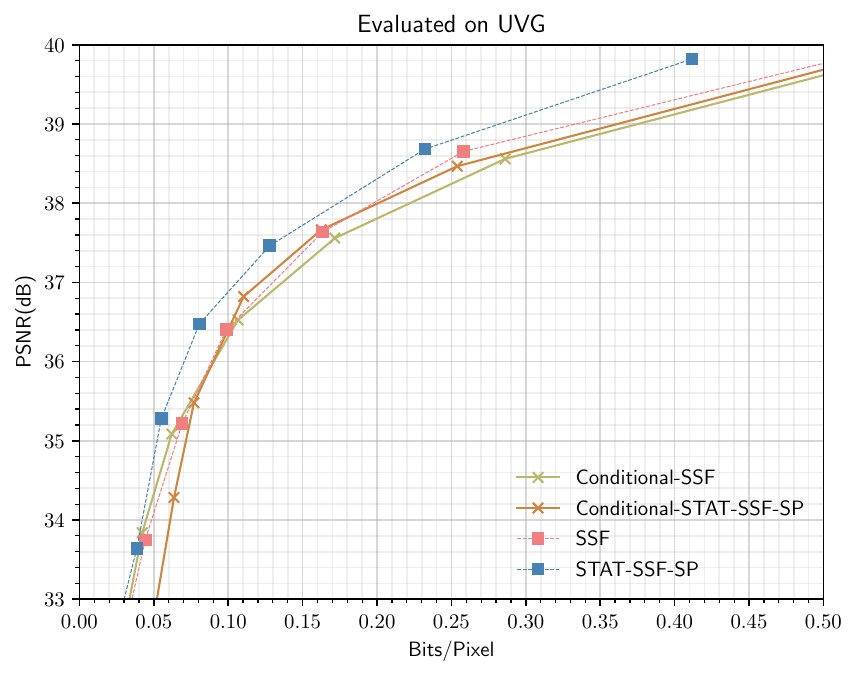}
        \caption{}
        \label{fig:vbr}
    \end{subfigure}
    \caption{
    \C{(a) Ablation study of STAT-SSF-SP, examining the effect of two proposed components, STAT (stochastic
temporal autoregressive transform) and SP (structured prior), with R-D results evaluated on the UVG dataset. Compared to STAT-SSF-SP, SSF-SP lacks the learned elementwise scaling transform in STAT (Sec.~\ref{sec:hybrid-method-stat}), STAT-SSF lacks the structured prior, while SSF \cite{agustsson2020scale} lacks both components. See discussion in Sec.~\ref{sec:base-results}. (b) Comparison of the Rate-Distortion performance between variable-bitrate models and non-variable-bitrate models}}
    \vspace{-1em}
    \label{fig: ablations and comparison}
\end{figure*}

\subsection{Base Results}
\label{sec:base-results}
\C{First, we examine the performance of our proposed approach without introducing temporal conditioning in the prior. }
Figure \ref{fig:uvg-rd} compares our proposed models (\textbf{STAT-SSF} and \textbf{STAT-SSF-SP}) with classical codec HEVC and baseline neural codecs on the UVG test dataset. As can be seen from the rate-distortion results, our \textbf{STAT-SSF-SP} model is uniformly better than previously known neural codecs represented by SSF \cite{agustsson2020scale} and DVC \cite{lu2019dvc}. \textbf{STAT-SSF-SP} even outperforms HEVC(YUV) (using YUV420 video input and evaluating on RGB444, the setting is available in the Supplementary Material) at bitrates $\geq 0.07$ BPP.
As expected, our proposed \textbf{STAT} model improves over \textbf{TAT} \cite{yang2020autoregressive}, with the latter lacking stochasticity in the autoregressive transform compared to our proposed STAT and its variants.

Figure \ref{fig:uvg-rd} shows that the performance ranking on MCL\_JCV is similar to that on UVG, despite MCL\_JCV having more diverse and challenging (e.g., animated) content. 
We provide qualitative results in Figure \ref{fig:qulitative-scale} and \ref{fig:qualitative-reconstructions}, offering insight into the behavior of the proposed scaling transform and structured prior, as well as visual qualities of different methods.

Additionally, we observe that different implementations of scale space volume (Gaussian pyramid or Gaussian blur) also slightly influence the rate-distortion performance. While the Gaussian blur version seems to have more advantage at the high bitrate regime, the Gaussian pyramid performs better at mid-range bitrate, as shown in Figure \ref{fig:main-rd}. The Gaussian pyramid usually generates a blurrier high-scale image compared to Gaussian blur when the base scales are the same.

\C{Finally, we ablate on our proposed STAT-SSF-SP model. This allows us to study the performance contribution of each of our proposed components, stochastic temporal autoregressive transform (STAT) and structured prior (SP), in isolation.
Compared to the baseline SSF \cite{agustsson2020scale} model, STAT-SSF introduces an additional elementwise scaling transform, while SSF-SP introduces the structured prior (see discussions in Sec.~\ref{sec:base-stat-models}).
As shown in Figure \ref{fig:ablation}, \textbf{SSF-SP} provides a comparable or greater improvement than \textbf{STAT-SSF}, relative to \textbf{SSF}. The improvements from the two components are complementary, and combining them into \textbf{STAT-SSF-SP} yields even further improvement as seen in Figure \ref{fig:uvg-rd}.
}

\subsection{Temporal Prior Experiments}
We also conduct experiments to illustrate the rate-distortion performance of \textbf{SSF} \cite{agustsson2020scale} and our proposed models using different temporal priors. Refer to table \ref{tab:table-model} for naming conventions.

\begin{figure*}[h]
    \centering
    \begin{subfigure}[]{.48\linewidth}
        \includegraphics[width=\linewidth]{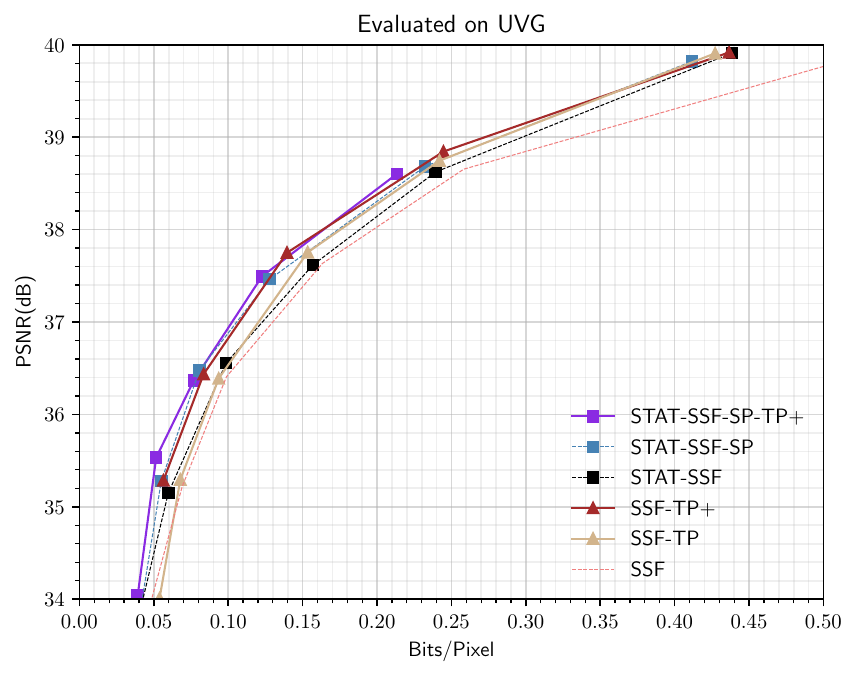}
        \caption{}
        \label{fig:uvg-ab}
    \end{subfigure}
    \begin{subfigure}[]{.48\linewidth}
        \includegraphics[width=\linewidth]{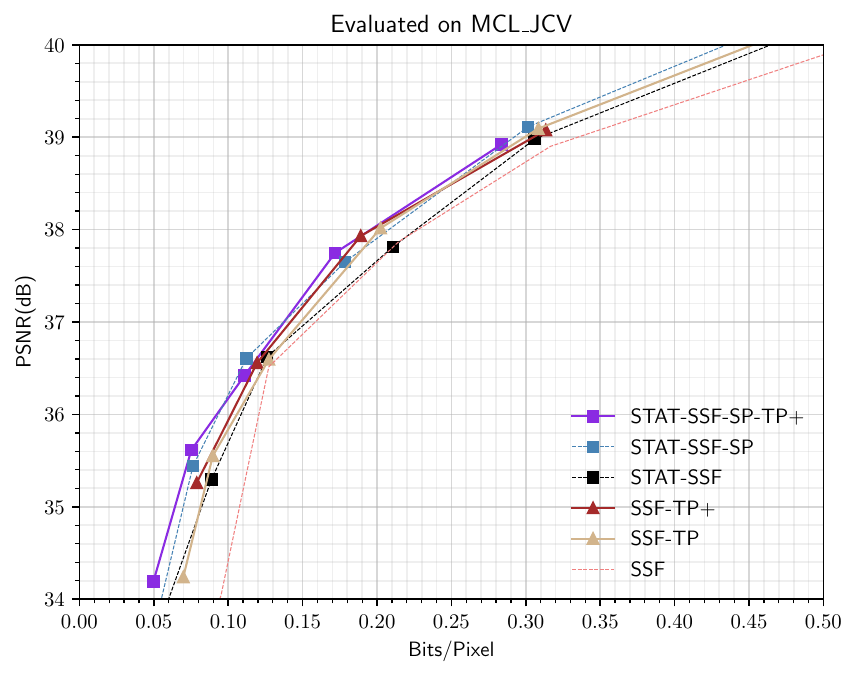}
        \caption{}
        \label{fig:mcl-ab}
    \end{subfigure}
    \caption{\textbf{Rate-Distortion Performance} of various models and ablations (see Table~\ref{tab:table-model}). Results are evaluated on \textbf{(a)} UVG and \textbf{(b)} MCL\_JCV datasets. The best results are obtained by models with deterministic temporal conditioning (TP+), while the improvement from conditioning on previous latents (TP) is less. }
    \label{fig:ab}
\end{figure*}

Figure \ref{fig:ab} shows the RD curve of each ablation model. We see that performing temporal conditioning through the prior does indeed improve performance, both for \textbf{STAT} and \textbf{SSF}. We see that temporally conditioning on the previous reconstruction (\textbf{SSF-TP+}) outperforms temporally conditioning on the previous latent (\textbf{SSF-TP}) in almost all bitrate regimes across both datasets. Finally, \textbf{SSF-TP+} even performs comparably with \textbf{STAT-SSF-SP}, and the hybrid model \textbf{STAT-SSF-SP-TP+} offers slight improvement further.

However, we also observe that compared with the \textbf{SSF-TP+} or \textbf{SSF} model, \textbf{SSF-TP} shows fluctuating rate-distortion performance during evaluation if we train the same model with different random initializations. We hypothesize that this may be caused by either numerical instability or video length inconsistency between training and evaluation data. To the best of our knowledge, most proposed sequential generative or compression models with a temporal prior are trained with more than three consecutive frames \cite{marino2020improving, li2018disentangled, denton2018stochastic, liu2019learned, yang2020Learning, yang2020learning2}. In contrast, our model only uses three consecutive frames during training for better efficiency.

Fortunately, we can still avoid the issue by using ``$\beta$-annealing''. We first train a model at a high bitrate and then finetune the model to lower bitrates using the increasingly larger $\beta$ value. In our experiment, the initial model is trained with $\beta=1.5625\times 10^{-4}$, and the following models are finetuned with $\beta=\{3.125\times 10^{-4}, 6.25\times 10^{-4}, 1.25\times 10^{-3}, 2.5\times 10^{-3}, 5\times 10^{-3}\}$, respectively.

\subsection{Variable-bitrate Experiments}

In contrast to previous experiments that train separate models, one for each $\beta$ value, in this section, we show results of the proposed conditional autoencoder based variable-bitrate scheme. The conditioning factor $B$ is defined as a one-hot vector with length 7 and seven $\beta$ values are used to match each one-hot vector: $\{10^{-2}, 5\times 10^{-3}, 2.5\times 10^{-3}, 10^{-3}, 5\times 10^{-4}, 2.5\times 10^{-4}, 10^{-4}\}$. For each training \emph{datum}, the $(B, \beta)$ pair is selected randomly. We also observe that this sampling scheme results in better rate-distortion performance than sampling $(B, \beta)$ per \emph{batch}.

In Figure \ref{fig:vbr}, we compare the performance of the variable-bitrate models and models with separate $\beta$ values. We observe that \textbf{Conditional SSF} performs slightly worse than \textbf{SSF} at high bitrate. While \textbf{Conditional STAT-SSF-SP} consistently has worse performance than \textbf{STAT-SSF-SP}, it performs better than \textbf{Conditional SSF} at bitrate $\leq 0.08$ bpp. This result demonstrates that the performance gap between variable and non-variable-bitrate models could be amplified with a more complicated model, while the gap is barely noticeable for simpler models. The investigation of this phenomenon will be left to future research.

\section{Discussion}
\label{sec: discussion}

Generative modeling and compression share similar aims, respectively modeling and removing redundant structure in data. Accordingly, the nascent field of learned, neural compression holds the potential to drastically improve performance over conventional codecs. Perhaps nowhere is this more impactful than in the setting of high-resolution video, where capturing temporal redundancy can yield significant reductions in coding costs. Toward this end, we have drawn inspiration from a recently developed technique within deep generative models, temporal autoregressive flows \cite{marino2020improving}, which perform temporal predictive coding via affine transforms. By interpreting 
recent state-of-the-art neural compression methods via autoregressive flows,
we can generalize them to using more complex transforms, with corresponding improvements in performance. We have investigated such improvements on video compression benchmarks, along with extensions to the higher-level latent variable model: additional hierarchical priors, temporal priors, and variable bitrate compression. Of particular importance, our results \C{show that domain knowledge (in our case, next frame prediction via the computer vision technique of warping) remains valuable in video compression and can be more cost-effective than a neural-network-only approach.} %
They also shed light on the potential pitfalls in estimating higher-level temporal dynamics in learned compression models.

\C{The art of transform coding involves the effective design of both the transform and entropy model, and many possibilities remain in the domain of video compression. %
Unlike for images, where the convolutional neural network has been established as the default  transform for capturing spatial redundancy \cite{balle2020nonlinear}, video introduces additional temporal redundancy, and it remains to be seen what transform is most effective while staying within (often stringent) computation budgets. The design space is vast, 
and the approach of predictive coding combined with computer vision techniques has so far achieved the widest commercial success.
Our work shows one way of extending this approach inspired by normalizing flows, and opens up the possibility of compression with more general and potentially non-affine autoregressive transforms \cite{huang2018neural}.
Finally, recent work \cite{mentzer2022vct} has shown state-of-the-art video compression performance with a rich entropy model and without predictive coding in the pixel space. This approach essentially offloads the task of designing good transforms to that of modeling high-dimensional discrete distributions, which now can be tackled thanks to the rise of large-scale transformer models \cite{vaswani2017attention} and abundance of data, but remains computationally prohibitive. 
An interesting future research direction could therefore be to explore the complementary strengths between this and our approach for sequence decorrelation, which may lead to further advances in both rate-distortion performance and computational efficiency.
}

\ifCLASSOPTIONcompsoc
  \section*{Acknowledgements}
Yibo Yang acknowledges funding from the Hasso Plattner Research Center in Machine Learning and Data Science at UC Irvine. Stephan Mandt acknowledges support by the National Science Foundation under the NSF CAREER Award 2047418 and grants CNS-2003237 and IIS-2007719, the Department of Energy under Grant  DE-SC0022331, as well as gifts from Intel, Disney, and Qualcomm.
\else
  \section*{Acknowledgements}
Yibo Yang acknowledges funding from the Hasso Plattner Research Center in Machine Learning and Data Science at UC Irvine. Stephan Mandt acknowledges support by the National Science Foundation under the NSF CAREER Award 2047418 and grants CNS-2003237 and IIS-2007719, the Department of Energy under Grant  DE-SC0022331, as well as gifts from Intel, Disney, and Qualcomm.
\fi

\ifCLASSOPTIONcaptionsoff
  \newpage
\fi

\bibliographystyle{IEEEtran}
\bibliography{tpami}

\vspace{-1.5cm}
\begin{IEEEbiography}[{\includegraphics[width=1in,clip,keepaspectratio]{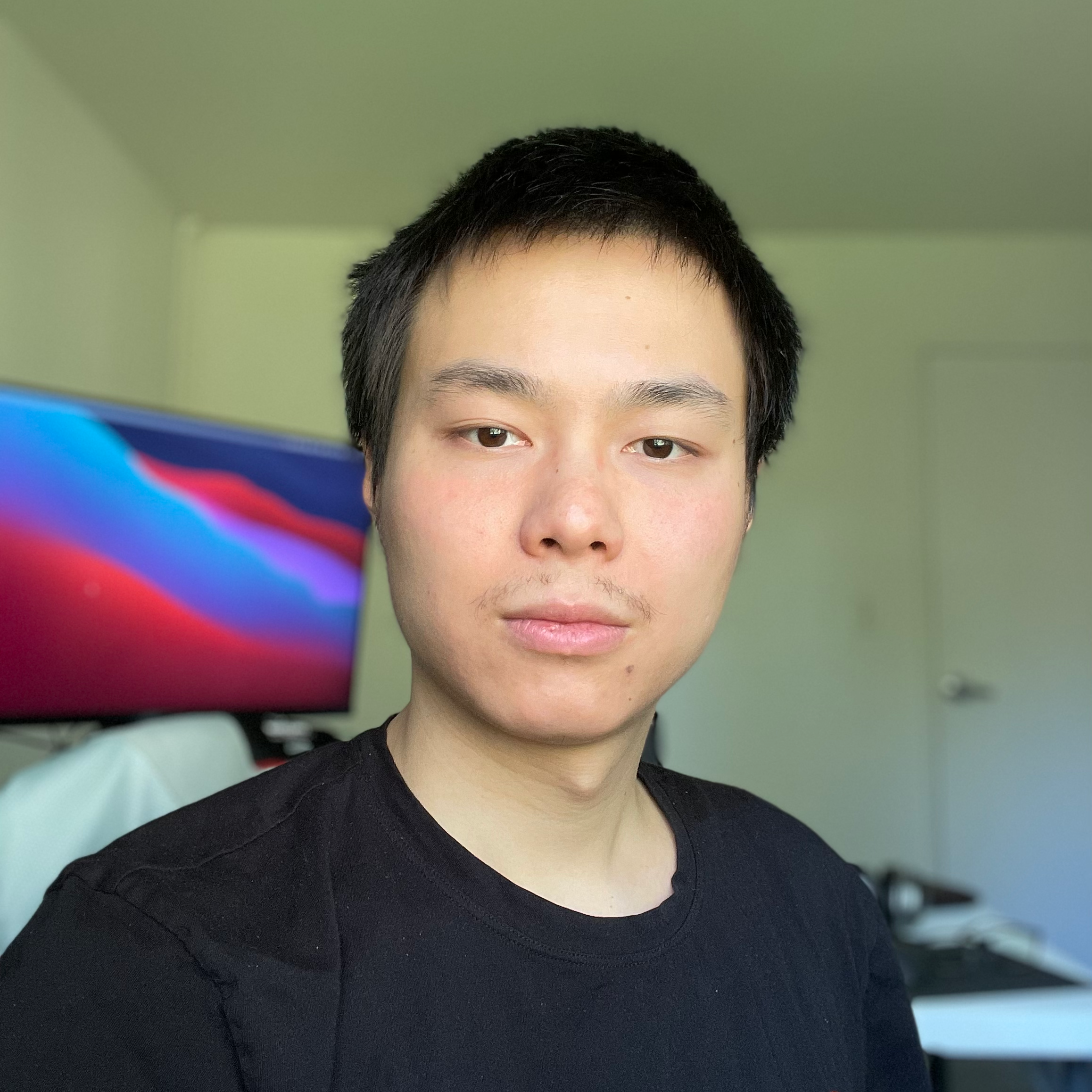}}] \newline \textbf{Ruihan Yang} received B.S. degree in Computer Science from NYU Shanghai in 2018. He is currently a Ph.D. student at the University of California, Irvine. From 2018 to 2019, his work focused on generative modeling of music, and had some experience on computational material science. He currently works on source compression with generative models.
\end{IEEEbiography}
\vspace{-1.5cm}
\begin{IEEEbiography}[{\includegraphics[width=1in,height=1.25in,clip,keepaspectratio]{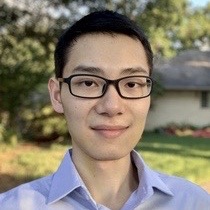}}] \newline \textbf{Yibo Yang} is a Ph.D. candidate at the University of California, Irvine. His research interests include probability theory, information theory, and their applications in machine learning. His recent work develops deep generative modeling approaches for data compression, and he has co-organized tutorials and workshops on the topic.
\end{IEEEbiography}
\vspace{-1.5cm}
\begin{IEEEbiography}[{\includegraphics[width=1in,height=1.5in,clip,keepaspectratio]{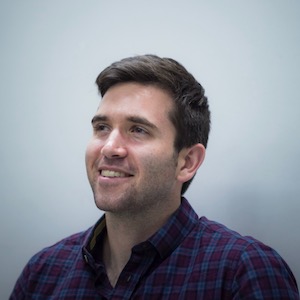}}] \newline \textbf{Joseph Marino} recently received a Ph.D. in computation and neural systems from the California Institute of Technology (Caltech). His work focuses on probabilistic models and inference algorithms as they relate to deep generative modeling, reinforcement learning, and theoretical neuroscience. He is currently employed as a research scientist at DeepMind.
\end{IEEEbiography}
\vspace{-1.5cm}
\begin{IEEEbiography}[{\includegraphics[width=1in,height=1.25in,clip,keepaspectratio]{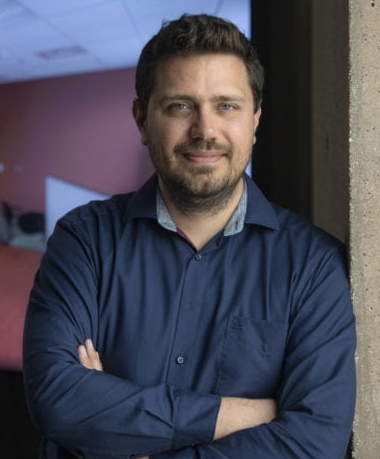}}]{Stephan Mandt} is an Associate Professor of Computer Science and Statistics at the University of California, Irvine. From 2016 until 2018, he was a Senior Researcher and Head of the statistical machine learning group at Disney Research in Pittsburgh and Los Angeles. He held previous postdoctoral positions at Columbia University and Princeton University. Stephan holds a Ph.D. in Theoretical Physics from the University of Cologne, where he received the German National Merit Scholarship. He is furthermore a recipient of the NSF CAREER Award, the UCI ICS Mid-Career Excellence in Research Award, the German Research Foundation's Mercator Fellowship, a Kavli Fellow of the U.S. National Academy of Sciences, a member of the ELLIS Society, and a former visiting researcher at Google Brain. Stephan is an Action Editor of the Journal of Machine Learning Research and of Transactions on Machine Learning Research. His research is currently supported by NSF, DARPA, IARPA, DOE, Disney, Intel, and Qualcomm.
\end{IEEEbiography}

\newpage
\section{Supplementary Material for ``Insights from Generative Modeling for Neural Video Compression'' }

\subsection{YouTube-NT dataset}
We provide a new full-resolution Youtube dataset\footnote{\url{https://github.com/privateyoung/Youtube-NT}} to the community that links to training data through scripts. This dataset complements existing data such as Vimeo-90k \cite{xue2019video} and contributes to standardized testing of video compression methods on high-resolution content. 

\begin{figure}[h]
    \centering
    \includegraphics[width=.97\linewidth]{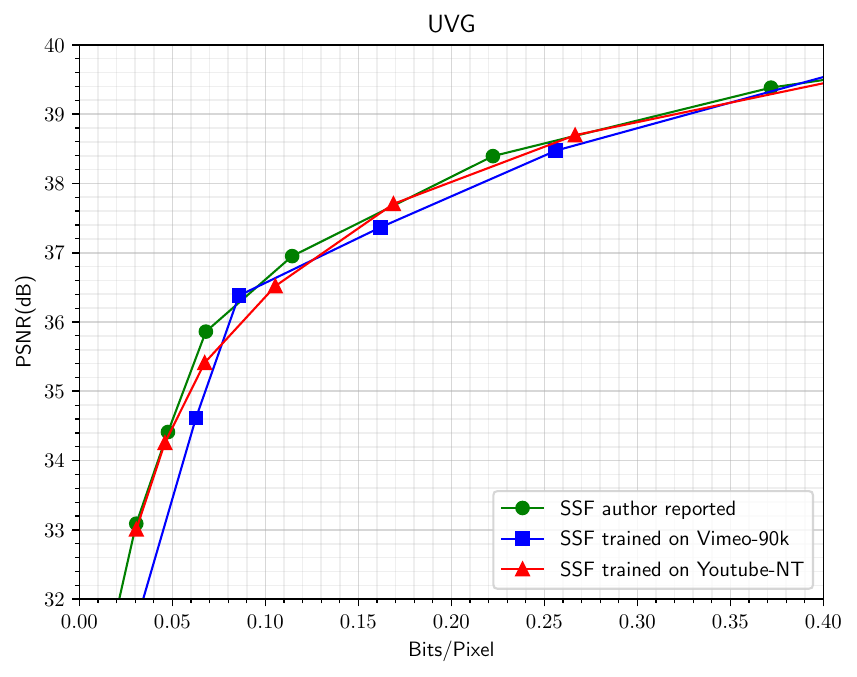}
    \caption{Comparison of rate-distortion performance on various datasets.}
    \label{fig:dataset}
\end{figure}

\textbf{Data collection.}
We collected 8,000 nature videos and movie/video-game trailers from \url{youtube.com} and processed them into 300k high-resolution (720p) clips, which we refer to as YouTube-NT.
We release YouTube-NT in the form of customizable scripts, which can preprocess any video that satisfy the requirements to facilitate future compression research. 
Table \ref{tab:dataset} compares the current version of YouTube-NT with Vimeo-90k \cite{xue2019video} and Google's private training dataset  \cite{agustsson2020scale}. Figure \ref{fig:dataset} shows the evaluation results of the SSF model trained on each dataset.

\begin{table*}[ht]
\centering
\caption{\textbf{Overview of Training Datasets}.}
\resizebox{0.9\textwidth}{!}{
\begin{tabular}{|c|c|c|c|c|c|c|}
\hline
\textbf{Dataset name}          & \textbf{Clip length} & \textbf{Resolution} & \textbf{\# of clips} & \textbf{\# of videos} & \textbf{Public} & \textbf{Configurable}\\
\hline\hline
Vimeo-90k             & 7 frames              & 448x256    & 90,000          & 5,000    & \cmark      & \checkmark\kern-1.1ex\raisebox{.7ex}{\rotatebox[origin=c]{125}{--}}       \\
\hline
YouTube-NT (\textbf{ours})            & 6-10 frames            & 1280x720   & 300,000         & 8,000     & \cmark     & \cmark      \\
\hline
Agustsson 2020 et al. & 60 frames             & 1280x720   & 700,000         & 700,000     & \xmark     & \xmark\\
\hline
\multicolumn{7}{|l|}{\makecell[l]{\textbf{Remark}: In Youtube-NT, all the features are configurable with the scripts.}}\\
\hline
\end{tabular}
}
\label{tab:dataset}
\end{table*}

\textbf{Comparison to previous results.}
To quantify the effect of training data on performance, we compare the test performance (on UVG) of the SSF model trained on Vimeo-90k \cite{xue2019video} and YouTube-NT. We also provide the reported results from \cite{agustsson2020scale}, which is trained on a larger (and unreleased) dataset. As can be seen from the R-D curves in Figure \ref{fig:dataset}, training on YouTube-NT improves rate-distortion performance over Vimeo-90k, in many cases bridging the gap with the performance from the larger private training dataset of \cite{agustsson2020scale}. At a higher bitrate, the model trained on Vimeo-90k\cite{xue2019video} tends to have a similar performance to YouTube-NT. This is likely because YouTube-NT currently only covers 8000 videos, limiting the diversity of the short clips.

\subsection{Command for HEVC codec}
\label{app:hevc}
To avoid \texttt{FFmpeg} package taking the advantage of the input file color format (YUV420), we first need to dump the \texttt{video.yuv} file to a sequence of lossless \texttt{png} files:

\begin{verbatim}
    ffmpeg -i video.yuv -vsync 0 \
      video/%
\end{verbatim}

Then we use the default low-latency setting in \texttt{ffmpeg} to compress the dumped \texttt{png} sequences:

\begin{verbatim}
    ffmpeg -y -i video/%
    -preset medium -x265-params bframes=0 \
    -crf {crf} video.mkv
\end{verbatim}

where \texttt{crf} is the parameter for quality control. The compressed video is encoded by HEVC with RGB color space.

To get the result of HEVC (YUV420), we directly execute:
\begin{verbatim}
    ffmpeg -pix_fmt yuv420p -s 1920x1080 \
    -i video.yuv -c:v libx265 -crf {crf} \
    -x265-params bframes=0 video.mkv
\end{verbatim}

\subsection{Training schedule}
\label{app:train}
Training time is about four days on an NVIDIA Titan RTX. Similar to \cite{agustsson2020scale}, we use the Adam optimizer \cite{kingma2014adam}, training the models for 1,050,000 steps. The initial learning rate of 1e-4 is decayed to 1e-5 after 900,000 steps, and we increase the crop size to 384x384 for the last 50,000 steps. All models are optimized using MSE loss.

\subsection{Efficient scale-space-flow implementation}
\label{app:ssf}
\cite{agustsson2020scale} uses a simple implementation of scale-space flow by convolving the previous reconstructed frame $\hat x_{t-1}$ with a sequence of Gaussian kernel $\sigma^2 = \{0, \sigma_0^2, (2\sigma_0)^2, (4\sigma_0)^2, (8\sigma_0)^2, (16\sigma_0)^2\}$. However, this  leads to a large kernel size when $\sigma$ is large, which can be computationally expensive. For example, a Gaussian kernel with $\sigma^2=256$ usually requires kernel size 97x97 to avoid artifact (usually $kernel\_size=(6*\sigma+1)^2$). To alleviate the problem, we leverage an efficient version of Gaussian scale-space by using Gaussian pyramid with upsampling. In our implementation, we use $\sigma^2 = \{0, \sigma_0^2, \sigma_0^2+(2\sigma_0)^2, \sigma_0^2+(2\sigma_0)^2+(4\sigma_0)^2, \sigma_0^2+(2\sigma_0)^2+(4\sigma_0)^2+(8\sigma_0)^2, \sigma_0^2+(2\sigma_0)^2+(4\sigma_0)^2+(8\sigma_0)^2+(16\sigma_0)^2\}$, because by using Gaussian pyramid, we can always use Gaussian kernel with $\sigma=\sigma_0$ to consecutively blur and downsample the image to build a \textit{pyramid}. At the final step, we only need to upsample all the downsampled images to the original size to approximate a scale-space 3D tensor. Detailed algorithm is described in Algorithm \ref{alg}.

\subsection{Lower-level Architecture Diagrams}
\label{app:archi}
\begin{figure*}[ht]
    \centering
    \includegraphics[width=\linewidth]{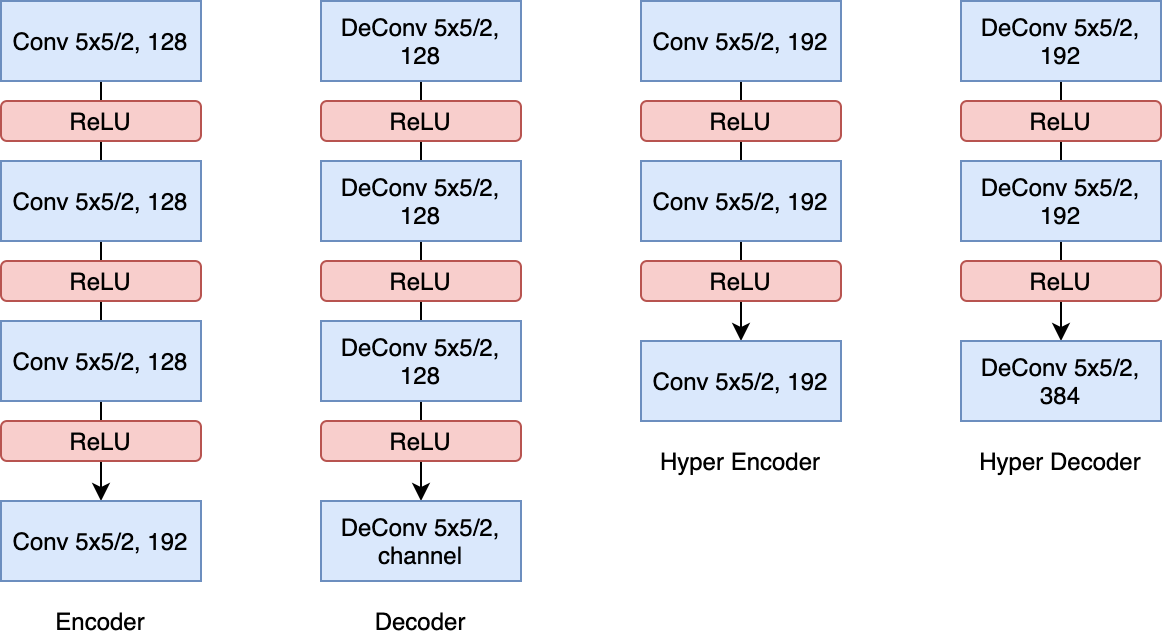}
    \caption{Backbone module architectures, where ``5x5/2, 128'' means 5x5 convolution kernel with stride 2; the number of filters is 128.}
    \label{fig:backbone}
\end{figure*}

\begin{figure*}[ht]
    \centering
    \includegraphics[width=\linewidth]{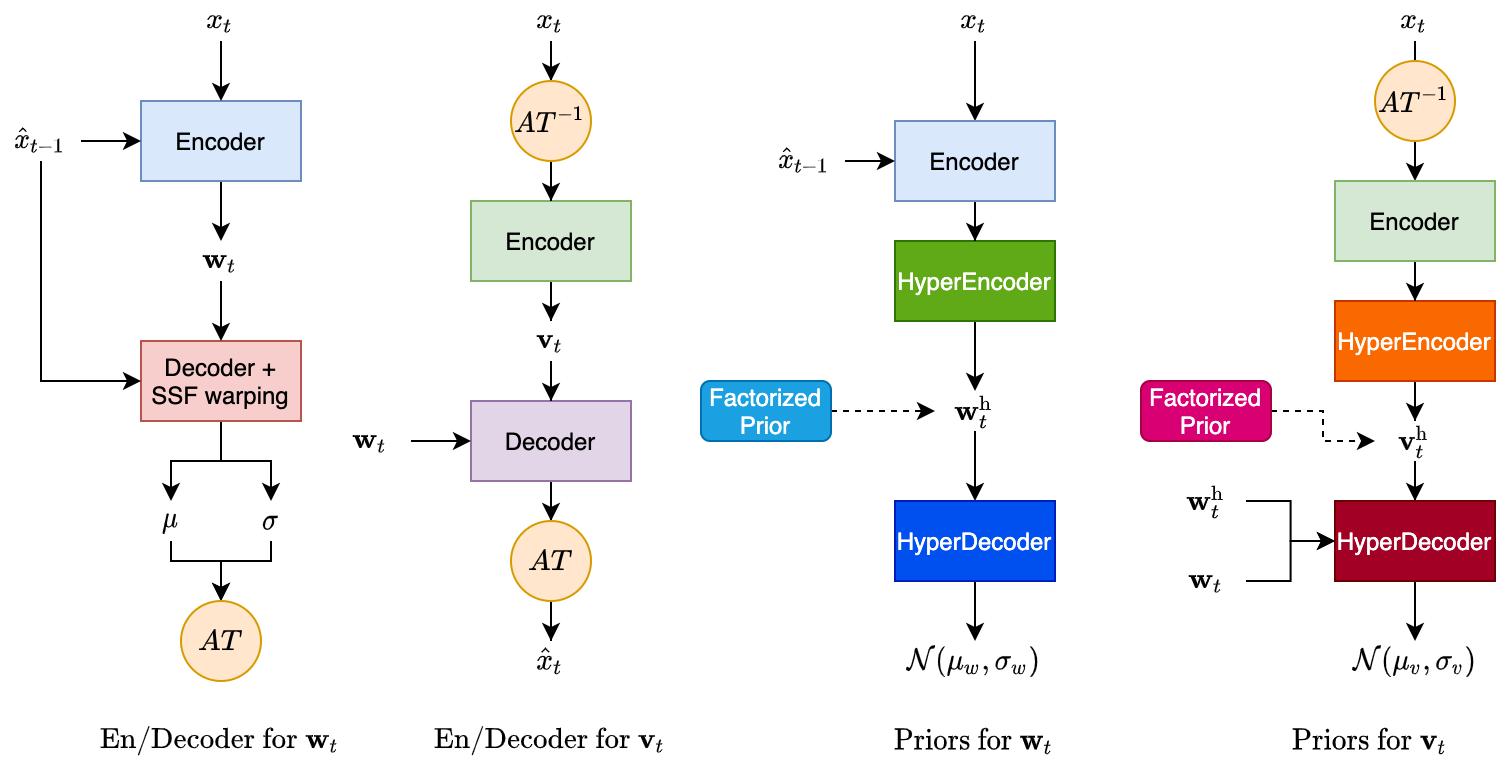}
    \caption{Computational flowchart for the proposed STAT-SSF-SP model. The left two subfigures show the decoder and encoder flowcharts for $\w_t$ and $\bv_t$, respectively, with ``AT'' denoting autoregressive transform. The right two subfigures show the prior distributions that are used for entropy coding $\w_t$ and $\bv_t$, respectively, with $p(\w_t, \w_t^\text{h}) = p(\w_t^\text{h}) p(\w_t | \w_t^\text{h})$, and $p(\bv_t, \bv_t^\text{h} | \w_t, \w_t^\text{h}) = p(\bv_t^\text{h}) p(\bv_t | \bv_t^\text{h}, \w_t, \w_t^\text{h})$, with $\w_t^\text{h}$ and $\bv_t^\text{h}$ denoting hyper latents (see \cite{agustsson2020scale} for a description of hyper-priors); 
    note that the priors in the SSF and STAT-SSF models (without the proposed structured prior) correspond to the special case where the HyperDecoder for $\bv_t$ does not receive $\w_t^{\text{h}}$ and $\w_t$ as inputs, so that the entropy model for $\bv_t$ is independent of $\w_t$: $p(\bv_t, \bv_t^\text{h}) = p(\bv_t^\text{h}) p(\bv_t| \bv_t^\text{h})$. }
    \label{fig:archi}
\end{figure*}

Figure \ref{fig:backbone} illustrates the low-level encoder, decoder and hyper-en/decoder modules used in our proposed STAT-SSF and STAT-SSF-SP models, as well as in the baseline TAT and SSF models, based on \cite{agustsson2020scale}. Figure \ref{fig:archi} shows the encoder-decoder flowchart for $\w_t$ and $\bv_t$ separately, as well as their corresponding entropy models (priors), in the STAT-SSF-SP model.

\end{document}